\documentclass[suppldata]{interact}

\usepackage{epstopdf}% To incorporate .eps illustrations using PDFLaTeX, etc.
\usepackage[caption=false]{subfig}% Support for small, `sub' figures and tables
\usepackage[natbibapa,nodoi]{apacite}% Citation support using apacite.sty. Commands using natbib.sty MUST be deactivated first!
\usepackage{mathrsfs}
\usepackage{graphicx}
\usepackage{setspace}
\usepackage{verbatim} 
\usepackage{natbib} 

{\end{list}}
\def\E{{\rm E}}

\def\bi{{\bf i}}

\def\bm{{\bf m}}

\def\bX{{\bf X}}
\def\bx{{\bf x}}
\def\bY{{\bf Y}}

\def\bz{{\bf z}}
\def\bbeta{\mbox{\boldmath $\beta$}}

\def\bmu{\mbox{\boldmath $\mu$}}

\def\b1{{\bf 1}}

\def\blot{\quad {$\vcenter{\vbox{\hrule height.4pt
             \hbox{\vrule width.4pt height.9ex \kern.9ex \vrule 
width.4pt}
             \hrule height.4pt}}$}}
\usepackage{latexsym}
\usepackage{amsmath}
\usepackage{amsfonts}
\usepackage{amssymb}
\usepackage{amsbsy}
\usepackage{amsthm}
\usepackage{graphicx}
\usepackage{colortbl}
\usepackage{comment}

\theoremstyle{plain}% Theorem-like structures provided by amsthm.sty
\newtheorem{theorem}{Theorem}[section]
\newtheorem{lemma}[theorem]{Lemma}

\theoremstyle{definition}

\theoremstyle{remark}

\begin{document}

\title{Statistical Uncertainty Analysis for Stochastic Simulation}
\author{
\name{Wei Xie\textsuperscript{1}, Barry L. Nelson\textsuperscript{2}, Russell R. Barton\textsuperscript{3}
\thanks{Corresponding author: Wei Xie (e-mail: w.xie@northeastern.edu).}}
\affil{\textsuperscript{1}Department of Mechanical and Industrial Engineering, Northeastern University 
\textsuperscript{2}Department of Industrial Engineering and Management Sciences, Northwestern University 
\textsuperscript{3}The Mary Jean and Frank P. Smeal College of Business Administration, Pennsylvania State University}
}
%\date{\vspace{-5ex}}
\date{}

\maketitle

\begin{abstract}
When we use simulation to evaluate the performance of a stochastic
system, the simulation often contains input distributions estimated
from real-world data; therefore, there is both simulation and input
uncertainty in the performance estimates.  Ignoring either source of
uncertainty underestimates the overall statistical error.  Simulation
uncertainty can be reduced by additional computation (e.g., more
replications).  Input uncertainty can be reduced by collecting more
real-world data, when feasible. 
This paper proposes an approach to quantify overall statistical
uncertainty when the simulation is driven by independent parametric input
distributions; specifically, we produce a confidence interval that
accounts for both simulation and input uncertainty by using a
metamodel-assisted bootstrapping approach. The input uncertainty is
measured via bootstrapping, an equation-based stochastic kriging
metamodel propagates the input uncertainty to the output mean, and
both simulation and metamodel uncertainty are derived using properties
of the metamodel. A variance decomposition is proposed to estimate the
relative contribution of input to overall uncertainty; this
information indicates whether the overall uncertainty can be
significantly reduced through additional simulation alone. Asymptotic
analysis provides theoretical support for our approach, while an
empirical study demonstrates that it has good finite-sample
performance.

\end{abstract}


\section{Introduction}

% set up the input uncertainty problem:

Stochastic simulation is used to characterize the behavior of complex
systems that are driven by random input processes. By ``input
process'' we mean a sequence of one or more random variables with a
fully specified joint distribution. In this paper we consider
independent and identically distributed (i.i.d.)\ input processes that are
themselves mutually independent, which means that the input processes
can be full specified by their individual marginal distributions.

The distributions of these input processes are often estimated from
real-world data.  Thus, a complete statistical characterization of
stochastic system performance requires quantifying both simulation and
input estimation errors.\footnote{As with any mathematical or computer
model, simulations are also subject to a host of non-statistical
errors in abstraction; these are not considered in this paper.} There
are robust methods that are adequate for quantifying simulation error
for many practical problems.  However, the impact of input estimation
error (which we call ``input uncertainty'') can overwhelm the
simulation error, as demonstrated in \cite{Barton_Schruben_2001};
ignoring it may lead to unfounded confidence in the simulation
assessment of system performance, which could be the basis for
critical and expensive decisions.  What we call input and simulation
uncertainty are also known as epistemic and aleatory uncertainty,
respectively; see for instance \cite{Kleijnen_2008}. Our method
accounts for both sources of error.

In this paper we address problems with parametric input distributions
that are mutually independent with parameters that are estimated from
a finite sample of real-world data. Of course, there exist practical
problems in which the input processes are not independent, and there
may also be significant uncertainty about the correct parametric
distribution as well as its parameter values.  Nevertheless, the case
of i.i.d.\ input processes represented by a parametric distribution is
prevalent in practice, making our contribution a useful step. We build
on \cite{barton_nelson_xie_2011}, which proposed a metamodel-assisted
bootstrapping approach that forms a confidence interval (CI) to
account for the impact of input uncertainty when estimating the
system's mean performance.  In that paper, bootstrap resampling of the
real-world data was used to approximate the input uncertainty, while a
metamodel predicted the simulation's mean response at different
parameter settings corresponding to bootstrap resampled data sets.

In a metamodel-assisted bootstrapping framework there are three types
of error: the input estimation error, the simulation estimation error
and the error in the metamodel itself. The latter two types of error
are not easily separated, so we call their combined effect ``metamodel
uncertainty." 

\cite{barton_nelson_xie_2011} showed that metamodel uncertainty can be
ignored when the simulation budget is not tight and an appropriate
type of metamodel and experiment design are used; they provided a
follow-up test to insure that the metamodel fit is close enough. In
this setting their method yields a valid CI. However, if the true mean
response surface is complex, especially for high-dimensional problems
(i.e., many input distributions), and the computational budget is tight, then
the impact of metamodel uncertainty can no longer be ignored without
underestimating the error, which manifests itself in a CI that is too
short. Computationally intensive stochastic simulations are the norm
for a number of fields: Spatial stochastic simulations, e.g., of oil
reservoirs, can take hours for a single run, and depend on many
stochastic parameters
(\citeauthor{Bangerth_2006}~\citeyear{Bangerth_2006},
\citeauthor{Wang_2012}~\citeyear{Wang_2012}).  Simulations of
semiconductor manufacturing
(\citeauthor{fowler2004grand}~\citeyear{fowler2004grand}) and
biological systems (\citeauthor{Ghosh_2011}~\citeyear{Ghosh_2011},
\citeauthor{Kastner_2002}~\citeyear{Kastner_2002}) can be similarly
intensive.

This paper is a significant enhancement of
\cite{barton_nelson_xie_2011}.  Here we propose an approach to form an
interval estimate that accounts for \textit{both} input and metamodel uncertainty in
estimating a stochastic system's mean performance.  When there is
little metamodel uncertainty the new method performs like
\cite{barton_nelson_xie_2011}, but it does not experience a
degradation in coverage when metamodel uncertainty is significant.

When the statistical error measured by our CI is too large for the
estimate to be useful, then the decision maker may want to know how
the error can be reduced. Our approach leads
naturally to a measure of the relative contribution of input to
overall uncertainty that indicates whether the error can be reduced by
an additional computational investment. 

The next section describes other approaches to attack the input
uncertainty problem and contrasts them with our method. This is
followed in Section~\ref{sec:problem_Description} by a formal
description of the problem of interest and a brief review of the
metamodel-assisted bootstrapping approach in Section~\ref{sec:MMABS}.
In Section 5 we provide an algorithm to build an interval estimator
accounting for both input and metamodel uncertainty, and give a method
to estimate their relative contributions. We then report results from
an empirical study of a difficult problem in Section 6 and conclude
the paper in Section 7. All proofs are in the Appendix.

\section{Background}
% put literature review here:

Various approaches to account for input uncertainty have been
proposed.  The Bayesian methods use the posterior distributions of the
inputs given the real-world data to quantify the input distribution
uncertainty, and the impact on the system mean is estimated by drawing
samples from these posterior distributions and running simulations at
each sample point (\citeauthor{Chick_2001}~\citeyear{Chick_2001};
\citeauthor{Chick_Ng_2002}~\citeyear{Chick_Ng_2002};
\citeauthor{Zouaoui_Wilson_2003}~\citeyear{Zouaoui_Wilson_2003},
\citeyear{Zouaoui_Wilson_2004}).  This could be computationally
expensive when the time for each simulation run is significant because
simulations need to be run at a large number of posterior sample
points. In addition, for each input prior distribution we need to
derive a corresponding posterior distribution which might be
nonstandard and complex. 

%A computationally expensive approach such as Markov Chain Monte Carlo
%may be needed to draw samples from these posterior distributions. 

A second approach is based on direct bootstrapping; it quantifies the
impact of input uncertainty using bootstrap resampling of the input
data, and runs simulations at each
bootstrap resample point to estimate the impact on the system mean
(\citeauthor{Barton_Schruben_2001}~\citeyear{Barton_Schruben_1993},
\citeyear{Barton_Schruben_2001};
\citeauthor{Barton_2007}~\citeyear{Barton_2007};
\citeauthor{Cheng_Holland_1997}~\citeyear{Cheng_Holland_1997}).
Compared with the Bayesian approach, the direct bootstrap can be
adapted to any input process without additional analysis and it
is suitable for complex and nonstandard input distributions. However,
similar to the Bayesian approach, this method also runs simulations at
each resample point. Since the number of bootstrap resample points
\textcolor{black}{to construct a CI} is
recommended to be a few thousand, the direct
bootstrapping method is also computationally expensive. More subtly,
since the statistic that is bootstrapped is the random output of a
simulation it is not a smooth function of the input data; this
violates the asymptotic validity of the bootstrap. 

Differing from the previous two approaches that estimate the system
mean response at each sample point by running simulations, a third
approach introduces an equation-based metamodel of the mean response
(see \cite{Cheng_Holland_2004} and references therein).  Specifically,
it assumes that the parametric families of the inputs are known, uses
maximum likelihood estimators (MLEs) of the unknown parameters, and
represents input-parameter uncertainty by the large-sample normal
distribution of the MLEs.  This uncertainty is propagated to the
output mean by a linear function of the parameters that is based on a
Taylor series approximation. Since the metamodel can be constructed
using simulation results from a small number of runs, this method does
not need substantial computational effort. However, a metamodel based
on a locally linear approximation is only appropriate when there is a
large quantity of real-world data so that the MLEs locate in a small
neighborhood of the true parameters with high probability; it is not
suitable when the underlying response surface is highly non-linear and
only a modest quantity of real-world data are available.  In addition,
the asymptotic normal approximation for the input distribution
parameters can be poor with sample sizes encountered in some
applications. 

The metamodel-assisted bootstrapping approach introduced by
\cite{barton_nelson_xie_2011} addresses some of the shortcomings in
the prior work. Compared with \cite{Cheng_Holland_2004}, the bootstrap
provides a more accurate approximation of the input uncertainty than
the asymptotic normal distribution of the parameter estimates in many
situations \citep{Horowitz_2001}. Further, the use of a
general-form metamodel provides higher fidelity than a locally linear
approximation. Compared with Bayesian and direct bootstrap approaches,
the use of a metamodel reduces the impact of simulation error on the
accuracy of CIs and reduces the computational effort because it does
not run simulations at a large number of sampled or resampled points;
instead, an equation-based metamodel is constructed based on a
designed experiment at a smaller number of parameter settings. In
addition, employing a metamodel makes the bootstrap statistic a smooth
function of the input data so that the asymptotic validity concerns
faced by the direct bootstrap method disappear.  The numerical results
in \cite{barton_nelson_xie_2011} provide evidence that
metamodel-assisted bootstrapping is effective and superior to
competitors \textit{when there is little metamodel uncertainty},
motivating its extension in this paper to more general and complex
input-uncertainty problems in which the impact of metamodel
uncertainty can no longer be ignored. The end result is a robust
method for quantifying statistical uncertainty.

\section{Problem Description}
\label{sec:problem_Description}

\begin{figure*}[tb]
\vspace{0.2in}
{
\centering
\includegraphics[scale=0.3]{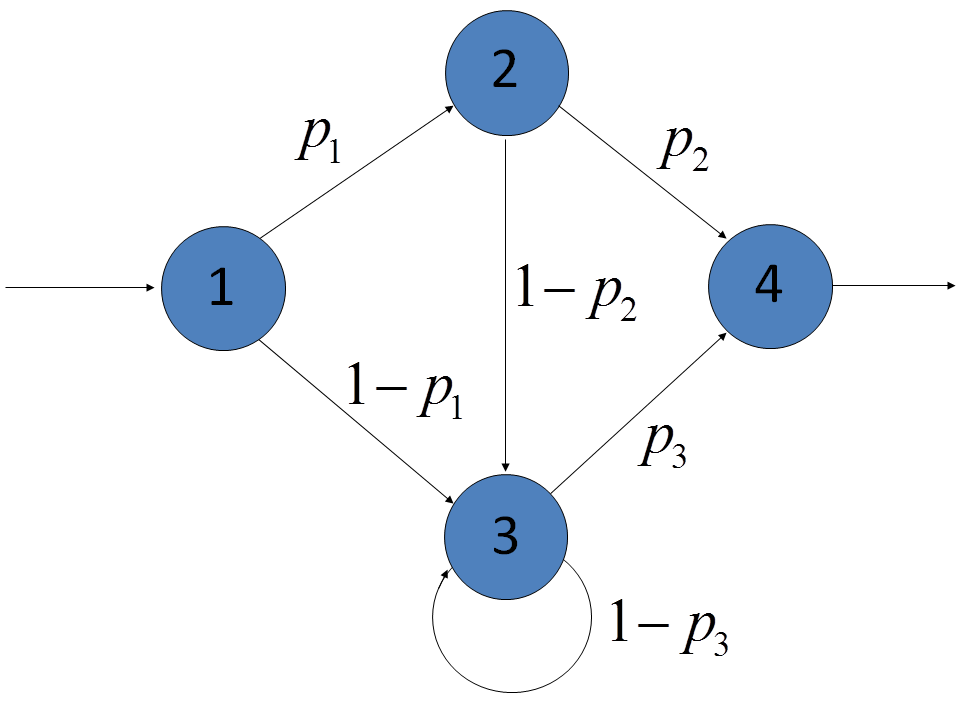}
\vspace{-0.2in}
\caption{Queueing network example.
\label{fig:Figure1_3}}
}
\end{figure*}

To make the description of the input uncertainty problem and our
solution to it easier to follow we will use the queueing network in
Figure~\ref{fig:Figure1_3} as an example and return to it in our
empirical study in Section~\ref{sec:empirical}. Consider estimating
the steady state expected number of customers in this network. The
interarrival times follow a gamma distribution, $A\sim
\mbox{gamma}(\alpha_A,\beta_A)$, and the service times at the $i$th
station also follow a gamma distribution,
$S_i\sim\mbox{gamma}(\alpha_{S_i},\beta_{S_i})$. Customers finishing
service at stations $1,2,3$ must make decisions about their next
station. These routing decisions follow Bernoulli distributions
$P_i\sim\mbox{Ber}(p_i), i=1,2,3$. The parameters of the input
distributions,
$\alpha_A,\beta_A$, $\{(\alpha_{S_i},\beta_{S_i}),i=1,2,3,4\}$ and
$\{p_i,i=1,2,3\}$ are all unknown and estimated from real-world data.
Notice that the inputs include both continuous and discrete
distributions. Our goal is to build a CI that covers the steady-state
expected number of customers in the system when the input parameters
assume their true but unknown values. We assume that at these ``true
values'' the system is in fact stable, and if we have enough
real-world data (which we may not) then the simulation with estimated
parameters will also be stable.

More generally, the stochastic simulation output is a function of
random numbers and $L$ independent input distributions
$F\equiv\{F_1,F_2,\ldots,F_L\}$. For notation simplification, we do
not explicitly include the random numbers. The output from the $j$th
replication of a simulation with input distribution $F$ can be written
as
\begin{equation}
Y_j(F)=\mu(F)+\epsilon_j(F) 
\label{eq.Y_F}
\end{equation}
where $\mu(F) = \E[Y_j(F)]$ denotes the unknown output mean and $\epsilon_j(F)$
represents the simulation error with mean zero. Notice that the
simulation output depends on the choice of input distributions. The
true ``correct'' input distributions, denoted by
$F^c\equiv\{F^c_1,F^c_2,\ldots,F^c_L\}$, are unknown and are estimated
from a finite sample of real-world data.  Our
goal is to quantify the impact of the statistical error by
finding a $(1-\alpha)100\%$ CI $[Q_L,Q_U]$ such that 
\begin{equation}
\mbox{Pr}\{\mu(F^c)\in [Q_L,Q_U]\} = 1-\alpha. 
\label{eq.CI_1}
\end{equation}
This is a relatively general statement of the problem
which could encompass multivariate input distributions (i.e., if one
or more of $F_\ell^c$ are multivariate) and also unknown distribution
families. 

However, in this paper we assume that the input
distributions are univariate, the families are known, but the
parameter values are not. 
%\begin{sloppypar} 
Therefore, the input distributions $F$ can be
completely characterized by the collection of parameters denoted by
$\bx$ with dimension $d$. We will define what we mean by ``parameters"
below. By abusing the notation, we rewrite Equation~(\ref{eq.Y_F}) as 
\begin{equation}
Y_j(\mathbf{x})=\mu(\mathbf{x})+\epsilon_j(\mathbf{x})
\label{eq.Y_x}
\end{equation}
where $\mathbf{x}\in \Psi$ and $\Psi \equiv\{\bx\in\Re^d{:} \mbox{ the
random variable } Y(\bx) \mbox{ is defined and $\mu(\bx)$ is
finite}\}$ denotes the region of interest.  The collection of true
parameters is denoted by $\bx_c$ and is assumed to be in the interior
of $\Psi$. We also assume $\mu(\bx)$ is continuous for $\bx\in\Psi$.
Thus, our goal can be restated as finding a $(1-\alpha)100\%$ CI such
that 
\begin{equation}
 \mbox{Pr}\{\mu(\bx_c)\in [Q_L,Q_U]\} = 1-\alpha. 
\label{eq.CI_2}
\end{equation}

Let $m_\ell$ denote the number of i.i.d.\ real-world observations
available from the $\ell$th input process
$\mathbf{Z}_{\ell,m_\ell}\equiv\left\{Z_{\ell,1},Z_{\ell,2},\ldots,Z_{\ell,m_\ell}
\right\}$ with $Z_{\ell,i}\stackrel{i.i.d}\sim F^c_\ell$,
$i=1,2,\ldots,m_\ell$. Let
$\mathbf{Z}_{\mathbf{m}}=\{\mathbf{Z}_{\ell,m_\ell},
\ell=1,2,\ldots,L\}$ be the collection of samples from all $L$ input
distributions in $F^c$, where $\mathbf{m}=(m_1,m_2,\ldots,m_L)$. The
real-world data are a particular realization of
$\mathbf{Z}_{\mathbf{m}}$, say $\mathbf{z}_{\mathbf{m}}^{(0)}$. Since
the unknown input distributions are estimated from
$\mathbf{z}_{\mathbf{m}}^{(0)}$, we assume the parameters are
functions of $\mathbf{Z}_{\mathbf{m}}$ denoted by
$\mathbf{X}_{\mathbf{m}}=\mathbf{X}(\mathbf{Z}_{\mathbf{m}})$.
Therefore, input uncertainty is fully captured by the sampling
distribution of $\mu(\mathbf{X}_{\mathbf{m}})$.

%, where from here on we treat $\mu(\cdot)$ as a function of the
%input-model parameters.

Since the underlying response surface $\mu(\cdot)$ is unknown, we
approximate it by a metamodel fitted to a set of stochastic simulation
outputs.  Let $\widehat{\mu}(\cdot)$ denote the metamodel. Fitting the
metamodel introduces a source of uncertainty in propagating the
sampling distribution of $\mathbf{X}_{\mathbf{m}}$ to the output mean:
metamodel uncertainty.  \textit{The contributions of this paper are to
construct an interval estimator that accounts for both input and metamodel uncertainty,
and to quantify the relative impact of each contributor. }

% ==============================================================
\section{Metamodel-Assisted Bootstrapping Framework}
\label{sec:MMABS}

\cite{barton_nelson_xie_2011} introduced the metamodel-assisted
bootstrapping approach to quantify input uncertainty.
We review it here.

The representation of the $L$ input distributions plays an important
role in the implementation of metamodel-assisted bootstrapping. Since
this paper focuses on problems with independent parametric
distributions having unknown parameters, $F$ can be uniquely
characterized by the corresponding collection of each distribution's
parameters or, in many cases, its moments. The $\ell$th input
distribution includes $h_\ell$ unknown parameters. Suppose that this
$h_\ell$-parameter distribution is uniquely specified by its first
(finite) $h_\ell$ moments, which is true for the distributions that
are most often used in stochastic simulation. The moments are chosen
as the independent variables for the metamodel because when they are
close, the corresponding distributions will be similar and therefore
generate similar outputs.  An extended argument for choosing moments
instead of the natural distribution parameters as independent
variables can be found in \cite{barton_nelson_xie_2011}. This
characterization will not work for all distributions, including some
so-called ``heavy-tailed'' distributions.

Let $\mathbf{x}_{[\ell]}$ denote an $h_\ell\times 1$ vector of the
first $h_\ell$ moments for the $\ell$th input distribution \textcolor{black}{and}
$d=\sum_{\ell=1}^L h_\ell$. By stacking $\bx_{[\ell]}$ with
$\ell=1,2,\ldots,L$ together, we have a $d\times 1$ dimensional
independent variable $\bx$.  Notice that $F$ is completely
characterized by the collection of moments $\bx$, and from here on
$\mu(\cdot)$ will be a function of distribution moments organized in
this way. Denote the true moments by $\bx_c$. 

For the queueing network in Figure~\ref{fig:Figure1_3}, there are
$L=8$ input distributions including arrivals, services at stations
$1,2,3,4$ and the three routing processes. Thus,
$h_1=h_2=h_3=h_4=h_5=2$ and $h_6=h_7=h_8=1$. The distributions for
the three routing processes can be completely specified by their
corresponding means. The distributions for the arrival and service
processes can be uniquely characterized by the corresponding first two
standardized moments: mean and standard deviation. To avoid a scaling
effect, the standard deviation is used instead of the second raw moment.
For the $\ell$th process, let $\tau_\ell$ denote the first moment
(mean) and $\varsigma_\ell$ denote the standard deviation. Then
$\bx=(\tau_1,\varsigma_1,\tau_2,\varsigma_2,\tau_3,\varsigma_3,\tau_4,\varsigma_4,
\tau_5,\varsigma_5,\tau_6,\tau_7,\tau_8)^\top$ with $d=13$.

The true moments $\bx_c$ are unknown and estimated based on a finite
sample $\mathbf{Z}_{\mathbf{m}}$ from $F^c$. As noted above,
$\mathbf{X}_{\mathbf{m}}$ is a $d\times 1$ dimensional moment
estimator that is a function of $\mathbf{Z}_{\mathbf{m}}$ written as
$\mathbf{X}_{\mathbf{m}}=\mathbf{X}(\mathbf{Z}_{\mathbf{m}})$.
Specifically,
$\mathbf{X}_{\ell,m_\ell}=\mathbf{X}_\ell(\mathbf{Z}_{\ell,m_\ell})$
and
$\mathbf{X}_{\mathbf{m}}^T=(\mathbf{X}_{1,m_1}^T,\mathbf{X}_{2,m_2}^T,\ldots,
\mathbf{X}_{L,m_L}^T)$. Let $F_{\mathbf{X}_{\mathbf{m}}}^c$ represent
the true, unknown distribution of $\mathbf{X}_{\mathbf{m}}$. Then
``input uncertainty'' refers to the distribution of
$\mu(\mathbf{X}_{\mathbf{m}})$ with $\mathbf{X}_{\mathbf{m}}\sim
F_{\mathbf{X}_{\mathbf{m}}}^c$.  Given a finite sample of real-world
data $\mathbf{z}_{\mathbf{m}}^{(0)}$, \cite{barton_nelson_xie_2011}
used bootstrap resampling to approximate
$F_{\mathbf{X}_{\mathbf{m}}}^c$ and a metamodel to represent
$\mu(\bx)$.

\subsection{Bootstrap Resampling}

We use distribution-free bootstrapping, meaning that we resample from
the empirical distribution of the data rather than a fitted parametric
distribution.  \textcolor{black}{Under some regularity conditions, the
bootstrap can provide an asymptotically consistent approximation for
the sampling distribution of a moment estimator \citep{Shao_1995}.
For the general performance of the bootstrap in representing the
sampling distribution of an estimator, see \cite{Horowitz_2001}.}

Implementation of the bootstrap in
metamodel-assisted bootstrapping is as follows. 
\begin{enumerate}

\item Draw $m_\ell$ samples with replacement from
$\mathbf{z}_{\ell,m_\ell}^{(0)}$, denoted by
$\mathbf{Z}_{\ell,m_\ell}^{(1)}$, and calculate the corresponding
$h_\ell\times 1$ vector of bootstrap moment estimates denoted by
$\widehat{{\mathbf{X}}}_{\ell,m_\ell}^{(1)}={\bX}_\ell
(\mathbf{Z}_{\ell,m_\ell}^{(1)})$ for $\ell=1,2,\ldots,L$. Then stack
the
results for all $L$ processes to obtain a $d\times 1$ vector
$\widehat{{\mathbf{X}}}_{\mathbf{m}}^{(1)}$.

\item Repeat the previous step $B$ times to generate
$\widehat{{\mathbf{X}}}_{\mathbf{m}}^{(b)},b=1,2,\ldots,B$.

\end{enumerate}
The bootstrap resampled moments are drawn from the bootstrap
distribution denoted by
$\widehat{F}_{\mathbf{X}_\mathbf{m}}(\cdot|\mathbf{z}_{\mathbf{m}}^{(0)})$.

\textcolor{black}{The most straightforward approach to propagate input
uncertainty to the output mean is direct bootstrapping. Given a total
computational budget of $N$ simulation replications, the system mean
response at each of $B$ bootstrap resamples is estimated by the sample
mean of $n=\lfloor N/B\rfloor$ replications, denoted by
$\bar{Y}(\widehat{\bX}_{\mathbf{m}}^{(b)}) = \sum_{j=1}^n
Y_j(\widehat{\bX}_\bm^{(b)}), b=1,2,\ldots,B$.  From these mean
estimates a CI is formed.  Direct bootstrapping consumes the total
simulation budget to estimate the simulation mean responses and to
form the CI.  Thus, for direct bootstrapping the impact of Monte Carlo
estimation error on the CI and the impact of simulation estimation
error on the mean responses are not separable.}

\cite{barton_nelson_xie_2011} assume that there is enough
computational budget available so that the difference between
$\widehat{\mu}(\bx)$ and $\mu(\bx)$ can be ignored; i.e.\
$\widehat{\mu}(\bx)\approx\mu(\bx)$.  Let
$\widehat{\mu}_b\equiv\widehat{\mu}(\widehat{\mathbf{X}}_{\mathbf{m}}^{(b)})$
for $b=1,2,\ldots,B$.  \cite{barton_nelson_xie_2011} quantified input
uncertainty by a $(1-\alpha)100\%$ bootstrap percentile interval
induced by the sorted responses $[Q^*_L,Q^*_U]=[\widehat{\mu}_{(\lceil
\frac{\alpha}{2}B\rceil)},\widehat{\mu}_{(\lceil(1-\frac{\alpha}{2})B\rceil)}]$,
where $\widehat{\mu}_{(i)}$ is the $i$th smallest value of
$\widehat{\mu}_1, \widehat{\mu}_2,\ldots, \widehat{\mu}_B$.  Here, the
superscript ``*" indicates that the input distribution is approximated
with the bootstrap. This interval provides asymptotically correct
coverage when all input distributions meet certain moment conditions
and the metamodel is continuously differentiable with nonzero gradient
in a neighborhood of $\mathbf{x}_c$ \citep{barton_nelson_xie_2011}.
Specifically, they establish the iterated limit
\[
\lim_{m\rightarrow \infty}\lim_{B\rightarrow
\infty}\mbox{Pr}\{\mu(\bx_c)\in [Q^*_L,Q^*_U]\}=1-\alpha 
\]
where as $m\rightarrow \infty$ we have $m_\ell/m \rightarrow 1$, for
$\ell=1,2,\ldots,L$.

However, with a tight computational budget for building the metamodel
we cannot guarantee that $\widehat{\mu}(\bx)\approx\mu(\bx)$ holds for
arbitrarily complex systems especially for problems with many input
distributions. Thus, we desire an interval estimator that accounts for
both input and metamodel uncertainty.  Stochastic kriging (SK),
introduced by \cite{ankenman_nelson_staum_2010}, facilitates this.  SK
is flexible.  Unlike the locally linear approximation in
\cite{Cheng_Holland_2004}, it does not require any strong assumption
about the form of the underlying true response surface $\mu(\cdot)$.
Based on our previous study \citep{xie_nelson_staum_2010}, a SK
metamodel fit to a small number of simulation runs can provide good
global predictions and also a characterization of metamodel
uncertainty for a wide variety of examples. The characterization of
metamodel uncertainty is a key contribution of the new method
presented here.

\textcolor{black}{\textit{Compared with direct bootstrapping,
metamodel-assisted bootstrapping separates the choice of $B$ from the
budget $N$ and reduces the influence of simulation estimation error.}
Instead doing simulations at $B$ samples from the bootstrap, we run
simulations at well-chosen design points and build an equation-based
metamodel $\widehat{\mu}(\bx)$ to predict the mean response at
different input distributions represented by bootstrap resampled
moments. Notice that once we have the metamodel, we can use any $B$ we
want to control the Monte Carlo estimation error of the interval
estimator, even $B>N$. Further, the metamodel can efficiently use the
computational budget to reduce the uncertainty introduced when
propagating the input uncertainty to the output mean.}

\subsection{Stochastic Kriging Metamodel}

Kriging is a widely used interpolation method. Since the outputs from
stochastic simulations include simulation variability that often
changes significantly across the design space, SK was introduced to
distinguish the uncertainty about the response surface from the
simulation error inherent in stochastic simulation output.  This
section provides a brief review of SK.

Suppose that the underlying true (but unknown) response surface can be
thought of as a realization of a stationary Gaussian Process (GP).
This fiction has been shown to provide a very useful framework for
quantifying uncertainty about the unknown surface implied by a
deterministic computer experiment and has been successfully employed
in a wide variety of applications (see, for instance,
\cite{Santer_2003}).  SK extends this framework to include the
variability of the simulation output $Y$ using the model 
\begin{equation}
\label{eq:sk}
Y_j(\mathbf{x})=\mathbf{\beta}_0+W(\mathbf{x})+\epsilon_j(\mathbf{x}).
\end{equation}
The independent variable $\bx$ is interpreted as a location in space.
In this paper, $\bx$ denotes a $d\times 1$ vector of moments that
uniquely characterize the input distributions.  The variation in the
simulation output over the $\mathbf{x}$ space is divided into
extrinsic (response-surface) uncertainty $W(\mathbf{x})$ and intrinsic
(simulation output) uncertainty $\epsilon_j(\mathbf{x})$.  The term
``intrinsic" uncertainty refers to the variability
inherent in the sampling that generates stochastic simulation outputs
and ``extrinsic" uncertainty refers to our lack of knowledge about the
response surface. 

SK uses a mean-zero, second-order stationary GP $W(\bx)$ to account
for the spatial dependence of the response surface. Thus, the
uncertainty about the true response surface $\mu(\bx)$ is represented
by a GP $M(\bx)\equiv\mathbf{\beta}_0+W(\mathbf{x})$ (note that
$\beta_0$ can be replaced by a more general trend term
$\mathbf{f}(\bx)^\top \bbeta$ without affecting our method). For many,
but not all, simulation settings the output is an average of a large
number of more basic outputs, so a normal approximation can be
applied: $\epsilon(\bx)\sim \mbox{N}(0,\sigma^2_{\epsilon}(\bx))$.  Of
course, normality of the simulation output will not always hold, but
could be empirically tested if there is a concern.

In SK, the covariance between $W(\mathbf{x})$ and
$W(\mathbf{x}^\prime)$ quantifies how knowledge of the surface at some
design points affects the prediction of the surface. A parametric form
of the spatial covariance, denoted by
$\Sigma(\mathbf{x},\mathbf{x}^\prime)=\mbox{Cov}[W(\mathbf{x}),W(\mathbf{x}^\prime)]
=\tau^2r(\mathbf{x}-\mathbf{x}^\prime)$, is typically assumed where
$\tau^2$ denotes the variance and $r(\cdot)$ is a correlation function
that depends only on the distance $\mathbf{x}-\mathbf{x}^\prime$.
Based on our previous study \citep{xie_nelson_staum_2010}, we use the
product-form Gaussian correlation function 
\begin{equation}
\label{eq:gauss.correlation}
  r(\mathbf{x}-\mathbf{x}^\prime)=
\exp \bigg(-\sum_{j=1}^d \theta_j(x_j-x^\prime_j)^2 \bigg)
\nonumber
\end{equation}
for the empirical evaluation in Section~6; however, our results do not
require it. Let
$\pmb{\theta}=(\theta_1,\theta_2,\ldots,\theta_d)$ represent the
correlation parameters; for different correlation functions the
dimension of $\pmb{\theta}$ could change.  In any event, $M(\bx)$ can
be represented by a Gaussian process $ M(\bx)\sim
\mathrm{GP}(\mathbf{\beta}_0,\tau^2 r(\mathbf{x}-\mathbf{x}^\prime)).$

\begin{sloppypar}
To reduce the uncertainty about $\mu(\bx)$ we choose an experiment
design consisting of pairs $\mathcal{D}
\equiv\{(\bx_i,n_i),i=1,2,\ldots,k\}$ at which to run simulations and
collect observations, where $(\bx_i,n_i)$ denotes the location and the
number of replications, respectively, at the $i$th design point. The
design that we recommend is described in more detail in the Appendix,
but it is not the only design that could be effective. The
simulation outputs at $\mathcal{D}$ are $\mathbf{Y}_\mathcal{D}\equiv
\left\{
(Y_1(\mathbf{x}_i),Y_2(\mathbf{x}_i),\ldots,Y_{n_i}(\mathbf{x}_i));
i=1,2,\ldots,k \right\}$ and the sample mean at design point $\bx_i$
is $\bar{Y}(\mathbf{x}_i)=\sum_{j=1}^{n_i}Y_j(\mathbf{x}_i)/n_i$.  Let
the sample means at all $k$ design points be
$\bar{\mathbf{Y}}_\mathcal{D}=(\bar{Y}(\mathbf{x}_1),\bar{Y}(\mathbf{x}_2),
\ldots,\bar{Y}(\mathbf{x}_k))^T$.
Since the use of common random numbers is detrimental to prediction
(as opposed to optimization; see \cite{chen_ankenman_nelson_2010}),
the simulations at different design
points are independent and the variance of
$\bar{\mathbf{Y}}_\mathcal{D}$ is represented by a $k\times k$
diagonal matrix
$C=\mbox{diag}\left\{\sigma^2_{\epsilon}(\bx_1)/n_1,\sigma^2_{\epsilon}(\bx_2)/n_2,
\ldots,\sigma^2_{\epsilon}(\bx_k)/n_k \right\}$. 
\end{sloppypar}

Let $\Sigma$ be the $k\times k$ spatial covariance matrix of the
design points and let $\Sigma(\bx,\cdot)$ be the $k\times 1$ spatial
covariance vector between each design point and a fixed prediction
point $\bx$.  If the parameters $(\tau^2,\pmb{\theta}, C)$ are known,
then the metamodel uncertainty can be characterized by a refined GP
$M_p(\bx)$ that denotes the conditional distribution of $M(\bx)$ given
all simulation outputs,
\begin{equation}
M_p(\bx)\sim \mathrm{GP}(m_{p}(\bx),\sigma^2_{p}(\bx))
\label{eq.posterior}
\end{equation}
where $m_p(\cdot)$ is the minimum mean squared error (MSE) linear
unbiased predictor 
\begin{equation}
m_{p}(\bx)=\widehat{\beta}_0
+\Sigma(\bx,\cdot)^\top(\Sigma+C)^{-1}
(\bar{\mathbf{Y}}_\mathcal{D}-\widehat{\beta}_0\cdot 1_{k\times 1}),
\label{eq.predictor1}
\end{equation}
and the corresponding variance is 
\begin{equation}
\label{eq.MSE1}
\sigma^2_{p}(\bx) =
\tau^2-\Sigma(\bx,\cdot)^\top(\Sigma+C)^{-1}\Sigma(\bx,\cdot)   
+\mathbf{\eta}^\top[1_{k\times 1}^\top(\Sigma+C)^{-1}1_{k\times 1} ]^{-1}\mathbf{\eta}  
\end{equation}
where $\widehat{\beta}_0=[1_{k\times 1}^\top(\Sigma+C)^{-1}1_{k\times
1}]^{-1}1_{k\times 1}^\top(\Sigma+C)^{-1}\bar{\mathbf{Y}}_\mathcal{D}$
and $\mathbf{\eta}=1-1_{k\times
1}^\top(\Sigma+C)^{-1}\Sigma(\bx,\cdot)$
\citep{ankenman_nelson_staum_2010}. 
With the parameters $(\tau^2,\pmb{\theta},{C})$ known, $M_p(\bx)$
depends on the simulation outputs only through
$\bar{\mathbf{Y}}_\mathcal{D}$. Thus, $M_p(\bx)$ is a random
\textcolor{black}{function} having the conditional distribution of
$M(\bx)$ given $\bar{\mathbf{Y}}_\mathcal{D}$.  \textcolor{black}{Notice
that $\sigma^2_p(\mathbf{x})$ reflects both metamodel and simulation
error, including the constant term $\widehat{\beta_0}$, with the intrinsic
simulation sampling error affecting $\sigma^2_p(\mathbf{x})$ through
the matrix $C$. }

Since in reality the spatial correlation parameters $\tau^2$ and
$\pmb{\theta}$ are unknown, MLEs are typically
used for prediction \textcolor{black}{with the log-likelihood function  
\begin{equation}
\ell(\beta_0, \tau^2,\pmb{\theta}) = -\ln[(2\pi)^{k/2}]-\frac{1}{2}\ln[|\Sigma+C|]
-\frac{1}{2}(\bar{\mathbf{Y}}_{\mathcal{D}}-\beta_0\cdot 1_{k\times 1})^\top
[\Sigma+C]^{-1}(\bar{\mathbf{Y}}_{\mathcal{D}}-\beta_0\cdot 1_{k\times 1}) 
\nonumber
\label{eq.loglikelihood}
\end{equation}
where $\Sigma$ is a function of $\tau^2$ and $\pmb{\theta}$.} The
sample variance is used as an estimate for the simulation variance at
design points $C$. By plugging $(\widehat{\beta}_0, \widehat{\tau}^2,
\widehat{\pmb{\theta}}, \widehat{C})$ into Equations~(\ref{eq.predictor1}) and
(\ref{eq.MSE1}) we can obtain the estimated mean
$\widehat{m}_p(\mathbf{x})$ and variance
$\widehat{\sigma}_p^2(\mathbf{x})$. Thus, the metamodel we use is
$\widehat{\mu}(\bx)=\widehat{m}_p(\bx)$ with marginal variance
estimated by $\widehat{\sigma}^2_p(\bx)$.

\textcolor{black}{\cite{ankenman_nelson_staum_2010} demonstrate that
$\widehat{m}_p(\bx)$ is still an unbiased predictor even with the
plug-in estimator $\widehat{C}$, and further that the variance
inflation of $\sigma^2_p(\bx)$ caused by using $\widehat{C}$ is
typically small. We performed an empirical study whose results
indicate that if we use an adequate experiment design, such as the
one-stage space-filling design used in this paper, then the
performance of metamodel-assisted bootstrapping is also not sensitive
to the estimation error in $\widehat{\tau}^2$ and
$\widehat{\pmb{\theta}}$; see the Appendix.  However, it is known that
the estimator~(\ref{eq.MSE1}) with plug-in MLEs may sometimes
underestimate the prediction variance; see \cite{DenHertog_2006}.}

In the derivations that follow we will assume that the parameters
$(\tau^2,\pmb{\theta},C)$ are known. This is necessary (and common in
the kriging literature) because including the effect of parameter
estimation is mathematically intractable. To apply the methods in
practice (including our empirical study below), we form plug-in
estimators by inserting
$\widehat{\tau}^2,\widehat{\pmb{\theta}},\widehat{C}$.

\section{Confidence Interval and Variance Decomposition}

Our approach is to use metamodel-assisted bootstrapping to provide a
CI for the true mean performance. To be robust the CI should account
for both input and metamodel uncertainty. Since $m_p(\bx)$ is an
unbiased predictor under the Gaussian process assumption,
$\sigma^2_p(\bx)=0$ for all $\bx$ would imply that there is no
metamodel uncertainty due either to a finite number of design points
$\bx_i$ or finite number of replications $n_i$; that is, $m_p(\bx) =
\mu(\bx)$.  Unfortunately, with anything short of complete
information, there will always be some metamodel uncertainty; and if
the budget is tight relative to the complexity of the true response
surface, then the effect of metamodel uncertainty could be
substantial, resulting in significant undercoverage of the confidence
interval of \cite{barton_nelson_xie_2011} as we show in
Section~\ref{sec:empirical}. The new interval introduced here does not
suffer this degradation, and therefore is robust to the amount of
simulation effort that can be expended and can be recommended for
general use.

\textcolor{black}{The kriging literature is the foundation for our work;
see for instance \cite{Santer_2003}. Kriging provides inference about
the value of an unknown function $\mu(\cdot)$ at a fixed prediction
point $\bx_0$ where the function has not been evaluated based on
values of the function at a set of design points.  Kriging models
uncertainty about the function as a GP $M(\cdot)$ by assuming
$\mu(\cdot)$ is a realization of $M(\cdot)$.  An interval constructed
to cover the conditional distribution of $M(\bx_0)$ given the values
at the design points is often interpreted as a CI for $\mu(\bx_0)$
(e.g., \cite{Picheny_2010}). The success of this paradigm is not
because the function of interest is actually random---it is not---but
because in many problems the conditional GP appears to be a robust
characterization of the remaining response-surface uncertainty.}

%with the desired probability based on the conditional distribution of
%$M(\bx_0)$ 

\textcolor{black}{We adopt the kriging paradigm but with two key
differences: our prediction point $\bx_c$ is also unknown and must be
estimated from real-world data, and our function $\mu(\cdot)$ can only
be evaluated in the presence of stochastic simulation noise. Given the
simulation outputs $\bar{\mathbf{Y}}_{\mathcal{D}}$, the remaining
uncertainty about $\mu(\cdot)$ is characterized by the conditional GP
$M_p(\cdot)$. To account for the impact from both input and metamodel
uncertainty, we construct an interval $[C_L,C_U]$ covering
$M_p(\bx_c)$ with probability $(1-\alpha)100\%$; that is,
\begin{equation}
 \mbox{Pr}\{M_p(\bx_c)\in [C_L,C_U]\} = 1-\alpha. 
\label{eq.CI_3}
\end{equation} }
\textcolor{black}{Since the conditional coverage is $1-\alpha$, the
unconditional coverage of $M(\bx_c)$ is $1-\alpha$ as well. The
revised objective~(\ref{eq.CI_3}) is connected to our
objective~(\ref{eq.CI_2}) through the assumption that the function
$\mu(\cdot)$ is a realization of the GP $M(\cdot)$. A procedure that
delivers an interval satisfying~(\ref{eq.CI_3}) will be a good
approximation for a CI procedure satisfying~(\ref{eq.CI_2}) if
$M_p(\cdot)$ faithfully represents the remaining uncertainty about
$\mu(\cdot)$.  This is clearly an approximation because in any real
problem $\mu(\cdot)$ is a fixed function, therefore we refer to
$[C_L,C_U]$ as an approximation for the CI (ACI).}

\begin{comment}
Let $\mbox{P}_{\bar{\mathbf{Y}}_{\mathcal{D}}}$ denote the
distribution of $\bar{\mathbf{Y}}_{\mathcal{D}}$.  Since the interval
$[C_L,C_U]$ satisfies
$\Pr\{M(\bx_c)\in[C_L,C_U]|\bar{\mathbf{Y}}_{\mathcal{D}}\}=1-\alpha$,
it covers $M(\bx_c)$ with probability $(1-\alpha)$ by 
\begin{equation}
\int\Pr\{M(\bx_c)\in [C_L,C_U]|
\bar{\mathbf{Y}}_{\mathcal{D}}\}d\mbox{P}_{\bar{\mathbf{Y}}_{\mathcal{D}}}=1-\alpha. 
\nonumber
\end{equation}
\end{comment}

%Under our assumptions, this is an equivalent reformulation of the
%interval $[Q_L,Q_U]$ in Equation~(\ref{eq.CI_2}) in that it requires
%the CI to cover the metamodel uncertainty about the true response at
%$\bx_c$, and therefore also cover $\mu(\bx_c)$ with the desired
%probability. If $\sigma^2_p(\bx_c)=0$ then the two formulations
%coincide. 

In a practical setting, what is the next step if the
\textcolor{black}{interval $[C_L,C_U]$} is so wide that we
are uncomfortable making decisions based on estimates with that level
of error? We suggest gaining some sense of the relative
contribution from each source of uncertainty as a guide toward either
running more simulations or collecting more real-world input data or
both.  For many problems collecting additional input data is not
feasible or we would have done so already; in such cases knowing that
input uncertainty is substantial and cannot be reduced allows us to
exercise caution in how we use the simulation results. 

In this section, we first present a procedure to build
\textcolor{black}{an ACI} that satisfies Equation~(\ref{eq.CI_3})
asymptotically. The asymptotic consistency of this
\textcolor{black}{interval} is proved under the assumption that the true
response surface is a realization of a GP \textcolor{black}{with all
parameters known except $\beta_0$}. Next a variance decomposition is
proposed to measure the relative contribution of input uncertainty to
overall statistical uncertainty, and we study its asymptotic
properties as well. This is a measure of input uncertainty due to
\textit{all} input distributions.  A method for attributing the input
uncertainty to the $L$ distributions is provided by
\cite{SongNelson_2013}.  Finally, we address problems that can arise
when the system's mean performance fails to exist, or the system is
undefined, for some values of the sample moments, and explain why
metamodel-assisted bootstrapping tends to be tolerant of the former
situation and can be adjusted for the latter.

Assumptions that are needed for the asymptotic analysis are the
following: 

\vspace{12pt}
\noindent
\textbf{Assumptions:}
\begin{enumerate}

\item The $\ell$th input distribution is uniquely determined by its
first $h_\ell$ moments and it has finite first $4h_\ell$ moments for
$\ell=1,2,\ldots,L$.

\item We have i.i.d observations
$Z_{\ell,1}^{(0)},Z_{\ell,2}^{(0)},\ldots,Z_{\ell,m_\ell}^{(0)}$ from
the $\ell$th distribution for $\ell=1,2,\ldots,L$. As $m\rightarrow
\infty$, we have $m_\ell/m \rightarrow c_\ell$, $\ell=1,2,\ldots,L$,
for a constant $c_\ell>0$.

\item The $\epsilon_j(\mathbf{x})\stackrel{i.i.d.}\sim
\mbox{N}(0,\sigma^2_{\epsilon}(\mathbf{x}))$ for any $\bx$, and
$M(\mathbf{x})$ is a stationary, separable GP with a continuous correlation
function satisfying
\begin{equation}
1-r(\mathbf{x}-\mathbf{x}^\prime)\leq\frac{c}
{|\mbox{log}(\parallel\mathbf{x}-\mathbf{x}^\prime \parallel_2)|^{1+\gamma}}
\mbox{ for all } \parallel\mathbf{x}-\mathbf{x}^\prime \parallel_2\leq\delta 
\label{eq.corrCond}
\end{equation}
for some $c>0$, $\gamma>0$ and $\delta<1$, where
$\parallel\mathbf{x}-\mathbf{x}^\prime \parallel_2
=\sqrt{\sum_{j=1}^d({x}_j-{x}^\prime_j)^2}.$ \item The input processes
$Z_{\ell j}^{(0)}$, simulation noise $\epsilon_j(\bx)$ and GP
$M(\mathbf{x})$ are mutually independent and the bootstrap process is
independent of all of them.
\end{enumerate}
\vspace{12pt}

Assumptions~1--2 give sufficient conditions for the almost
sure (a.s.) consistency of bootstrap moment estimators
$\widehat{\mathbf{X}}_{\mathbf{m}}\stackrel{a.s.}\rightarrow\bx_c$ as
$m\rightarrow\infty$ (see Lemma~1 in the Appendix). Under
Assumption~3, a GP $M(\cdot)$ with a correlation function satisfying
Condition~(\ref{eq.corrCond}) has continuous sample paths almost
surely (\citeauthor{Adler_2010}~\citeyear{Adler_2010}, Theorem 3.4.1).
Condition~(\ref{eq.corrCond}) is satisfied by many correlation
functions used in practice, and in particular any power exponential
correlation function $r(\bx-\bx^\prime)= \exp\left(- \sum_{j=1}^d
\theta_j|x_j-x_j^\prime|^p \right)$ with $0<p\leq 2$ and $\theta_j>0$
\citep{Santer_2003}. Assumption~4 indicates that input data are
collected independently of the simulation model, and that our
uncertainty about the mean response surface as represented by $M(\bx)$
is independent of the stochastic simulation noise (although both can
depend on $\bx$).

\subsection{ACI Procedure}
\label{subSec:procedure}

Based on a hierarchical approach, we propose the following procedure
to build $(1-\alpha)100\%$ bootstrap percentile ACIs to
achieve~(\ref{eq.CI_3}):

\begin{enumerate}

\item Given real-world data $\mathbf{z}_\mathbf{m}^{(0)}$, choose
experiment design $\mathcal{D}=\{(\bx_i,n_i), i=1,2,\ldots, k\}$ as described in
the Appendix. 

\item Run simulations at design points to obtain outputs
$\mathbf{Y}_\mathcal{D}$. Compute the sample average $\bar{Y}(\bx_i)$
and sample variance $S^2(\bx_i)$ of the simulation outputs,
$i=1,2,\ldots, k$.  \textcolor{black}{Fit the SK metamodel parameters $(\beta_0, \tau^2,
\pmb{\theta}, C)$} to obtain $\widehat{m}_p(\bx)$ and
$\widehat{{\sigma}}_{p}^2(\mathbf{x})$ using $\left(\bar{Y}(\bx_i),
S^2(\bx_i), \bx_i\right)$, $i=1,2,\ldots, k$.

\item For $b = 1 \mbox{ to } B$

\begin{enumerate}

\item Generate bootstrap resample $\mathbf{Z}^{(b)}_{\mathbf{m}} 
\stackrel{i.i.d.}\sim \mathbf{z}^{(0)}_{\mathbf{m}}$ and compute sample
moments $\widehat{\mathbf{X}}_{\mathbf{m}}^{(b)}$.

\item Let $\widehat{\mu}_b\equiv\widehat{m}_p(\widehat{\mathbf{X}}_{\mathbf{m}}^{(b)})$.

\item Draw $\widehat{M}_b \sim \mbox{N}\left( \widehat{m}_p (
\widehat{\mathbf{X}}_{\mathbf{m}}^{(b)}),
\widehat{\sigma}_{p}^2(\widehat{\mathbf{X}}_{\mathbf{m}}^{(b)})
\right)$.  

\end{enumerate}

\item[] Next $b$

\item Report estimated CI and ACI, respectively,
\begin{eqnarray*}
\mbox{CI}_0 &\equiv& \left[
\widehat{\mu}_{(\lceil B\frac{\alpha}{2}
\rceil)},\widehat{\mu}_{(\lceil B(1-\frac{\alpha}{2})\rceil)} \right]
\\
\mbox{CI}_+ &\equiv& \left[
\widehat{M}_{(\lceil B\frac{\alpha}{2} \rceil)},\widehat{M}_{(\lceil
B(1-\frac{\alpha}{2})\rceil)}\right]
\end{eqnarray*}
where $\widehat{\mu}_{(1)} \le \widehat{\mu}_{(2)} \le \cdots \le
\widehat{\mu}_{(B)}$ and 
$\widehat{M}_{(1)} \le \widehat{M}_{(2)} \le \cdots \le
\widehat{M}_{(B)}$ are the sorted values. 
\end{enumerate}

In this procedure, Step~1 provides an experiment design to build a SK
metamodel, which is central to the metamodel-assisted bootstrapping
approach.  Since the input uncertainty is quantified with bootstrap
resampled moments, we want the metamodel to correctly predict the
responses at these points $\widehat{\mathbf{X}}_{\mathbf{m}}\sim
\widehat{F}_{\mathbf{X}_{\mathbf{m}}}(\cdot|\mathbf{z}_{\mathbf{m}}^{(0)})$.
Thus, the metamodel needs to be accurate and precise in a design space
that covers the ``most likely" bootstrap moment estimates, which can
be achieved by the experiment design proposed by
\cite{barton_nelson_xie_2011}. Their design is data-driven;
specifically, they first find the smallest ellipsoid denoted by $E$
that covers the most likely bootstrap resampled moments.  They then
generate a space-filling design that covers $E$. This design
\textcolor{black}{methodology, which is summarized in the Appendix}, yielded
accurate metamodels in the examples they studied. 

Based on the experiment design provided in Step~1, we run simulations
and construct a metamodel in Step~2 \textcolor{black}{by fitting $(\beta_0, \tau^2,
\pmb{\theta}, C)$}. Given the metamodel, we predict
the simulation's mean responses at different input settings
corresponding to bootstrap resampled moments and construct interval
estimators as shown in Step~3. Notice that Step~3(a) accounts for the
input uncertainty and Step~3(c) accounts for the input and metamodel
uncertainty.  Thus, this procedure provides two types of intervals:
\begin{itemize}

\item $\mbox{CI}_0$, proposed in \cite{barton_nelson_xie_2011},
returns an estimate of $[Q_L,Q_U]$ in Equation~(\ref{eq.CI_2}) by
assuming $\widehat{m}_p(\bx) = \mu(\bx)$; that is, it only accounts
for input uncertainty and will be in error if there is substantial
metamodel uncertainty.

\item  $\mbox{CI}_+$ returns an estimate of $[C_L,C_U]$ in
Equation~(\ref{eq.CI_3}). This ACI accounts for both input and
metamodel uncertainty.

\end{itemize}
As the metamodel uncertainty decreases, $\mbox{CI}_0$ and
$\mbox{CI}_+$ become closer and closer to each other.
Before evaluating the finite-sample performance of $\mbox{CI}_+$ in
Section~6, we establish its asymptotic consistency
\textcolor{black}{for objective~(\ref{eq.CI_3})}. 

\textcolor{black}{In Theorems~1--3
that follow, we replace $\widehat{\mu}_b$ and $\widehat{M}_b$ in
Steps~3(b)--(c) of the ACI
procedure with}
\[
\textcolor{black}{\mu_b \equiv m_p(\widehat{\bX}_{\bm}^{(b)})
\mbox{ and }}
M_b \sim \mbox{N}\left( m_p (
\widehat{\mathbf{X}}_{\mathbf{m}}^{(b)}),
\sigma_{p}^2(\widehat{\mathbf{X}}_{\mathbf{m}}^{(b)})
\right)
\]
recalling that $m_p(\cdot)$ and $\sigma_p^2(\cdot)$ are the marginal
mean and variance of the conditional GP $M_p(\cdot)$ \textcolor{black}{when $(\tau^2,
\pmb{\theta}, C)$ are known}.

\begin{theorem} \label{thm:consistency}
Suppose that Assumptions 1--4 hold.
Then the interval $[M_{(\lceil B\frac{\alpha}{2} \rceil)},
M_{(\lceil B(1-\frac{\alpha}{2})\rceil)}]$ is asymptotically
consistent, meaning the iterated limit
\begin{equation}
\lim_{m\rightarrow\infty} \lim_{B\rightarrow \infty} 
\Pr\{M_{(\lceil B\alpha/2 \rceil)} \leq M_p(\mathbf{x}_c) 
\leq M_{(\lceil B(1-\alpha/2)\rceil)} \} = 1-\alpha.
\label{eq.11_1}
\end{equation} 
\end{theorem}

In brief, under the assumption that $\mu(\bx)$ is a realization of a
GP, $M_p(\bx)$ characterizes the remaining metamodel uncertainty after
observing $\bar{{\mathbf{Y}}}_\mathcal{D}$. And since the input
uncertainty is asymptotically correctly quantified by the bootstrap
moment estimator $\widehat{\mathbf{X}}_{\mathbf{m}}$, the distribution
of $M_p(\widehat{\mathbf{X}}_{\mathbf{m}})$ accounts for both input
and metamodel uncertainty.
\textcolor{black}{Theorem~\ref{thm:consistency} shows that this interval
satisfies objective~(\ref{eq.CI_3}) asymptotically.  We are
particularly interested in situations when the simulation effort is
limited (\cite{barton_nelson_xie_2011} addressed the ample budget
case), so the consistency result in Theorem~\ref{thm:consistency} is
only with respect to the real-world data.} The detailed proof is
provided in the Appendix.

\textcolor{black}{In practice, including our empirical evaluation in
Section~\ref{sec:empirical}, $(\tau^2, \pmb{\theta}, C)$ must be estimated,
and the impact of parameter estimation (other than $\bx_c$ and
$\beta_0$) is not covered by Theorem~1. We address sensitivity to parameter
estimation in the Appendix.}

\begin{comment}
\textcolor{black}{Our goal is to find a CI that can quantify the impact
of input and simulation estimation error on the system performance
estimate and satisfy Equation~(\ref{eq.CI_2}). The revised
objective~(\ref{eq.CI_3}) is mainly used to facilitate analysis.
Therefore, the empirical test in Section~\ref{sec:empirical} evaluates
the finite sample performance about the probability of the derived
interval $\mbox{CI}_+$ covering $\mu(\bx_c)$ instead $M_p(\bx_c)$.  }
\end{comment}

%We use a hierarchical approach to generate $B$ samples for
%$M_p(\widehat{\mathbf{X}}_{\mathbf{m}})$ and construct the interval
%$[M_{(\lceil B\alpha/2 \rceil)}, M_{(\lceil B(1-\alpha/2)\rceil)}]$. 

\subsection{Variance Decomposition}
\label{subSec:varDecomp}

$\mbox{CI}_+$ accounts for input and metamodel uncertainty.  When the
width of $\mbox{CI}_+$ is too large to be useful, it is important to
know the relative contribution from each source. Since the total
output variability is the convolution of the input uncertainty and
simulation/metamodel uncertainty, it is hard to separate the effects
from these sources. To estimate the relative contributions
\cite{Zouaoui_Wilson_2003}, \cite{Ng_Chick_2006},
\cite{Ankenman_Nelson_2012} and \cite{SongNelson_2013} assume that the
simulation noise has a constant variance. In this section, we propose
a variance decomposition that does not require the homogeneity
assumption. 

Suppose that the parameters $(\tau^2,\pmb{\theta},C)$ are known, the
metamodel uncertainty can be characterized by a GP and the simulation
error follows a normal distribution. Then the metamodel
uncertainty, given the simulation result
$\bar{\mathbf{Y}}_\mathcal{D}$, is characterized by a GP $M_p(\bx)\sim
\mbox{N}(m_p(\mathbf{x}),\sigma^2_{p}(\mathbf{x}))$.  Conditional on
$\bar{\mathbf{Y}}_\mathcal{D}$, both $m_p(\mathbf{x})$ and
$\sigma^2_{p}(\mathbf{x})$ are fixed functions.  For notation
simplification, all of following derivations are conditional on the
simulation outputs $\bar{\mathbf{Y}}_\mathcal{D}$, but we will
suppress the ``$|\bar{\mathbf{Y}}_\mathcal{D}$''. 

The random variable $M_p({\mathbf{X}}_{\mathbf{m}})$ accounts for
input uncertainty through the sampling distribution of
$\mathbf{X}_{\mathbf{m}}$ and the metamodel uncertainty through the
random function $M_p(\cdot)$. 
To quantify the relative contribution of input and metamodel
uncertainty, we decompose the total variance of
$M_p({\mathbf{X}}_{\mathbf{m}})$ into two parts:
\textcolor{black}{
\begin{eqnarray}
\sigma^2_T &\equiv& \mbox{Var}[M_p(\mathbf{X}_{\mathbf{m}})] \nonumber \\[6pt]
&=&\mbox{E}\{\mbox{Var}[M_p(\mathbf{X}_{\mathbf{m}})|\mathbf{X}_{\mathbf{m}}]\}
+\mbox{Var}\{\mbox{E}[M_p(\mathbf{X}_{\mathbf{m}})|\mathbf{X}_{\mathbf{m}}]\} \nonumber\\[6pt] 
&=& \mbox{E}[\sigma^2_{p}(\mathbf{X}_{\mathbf{m}})]
+\mbox{Var}[m_p(\mathbf{X}_{\mathbf{m}})].
\label{eq.varDecomp}
\end{eqnarray}}
The term $\sigma^2_M\equiv
\mbox{E}[\sigma^2_{p}(\mathbf{X}_{\mathbf{m}})]$ is a measure of the
metamodel uncertainty: the expected metamodel variance weighted by the
density of moment estimator $\mathbf{X}_{\mathbf{m}}$.  This weighting
makes sense because the accuracy of the metamodel in regions with
higher density is more important for the estimation of system mean
performance. The term $\sigma^2_I\equiv
\mbox{Var}[m_p(\mathbf{X}_{\mathbf{m}})]$ is a measure of input
uncertainty when we replace the unknown true response surface
$\mu(\cdot)$ with its best linear unbiased estimate $m_p(\cdot)$. 

What is the contribution of each term to ACI coverage?  If the
metamodel uncertainty disappears (i.e., $\sigma^2_p(\cdot)=0$), then
$\sigma^2_M=0$, $\mbox{CI}_0$ and $\mbox{CI}_+$ coincide and they
provide asymptotically consistent coverage
\citep{barton_nelson_xie_2011}. Metamodel uncertainty is reduced by
simulation effort. On the other hand, as $m\rightarrow\infty$ (more
and more real-world input data),
$\mathbf{X}_{\mathbf{m}}\stackrel{a.s.}\rightarrow\bx_c$ and since
$m_p(\bx)$ is continuous we have $\sigma^2_I=0$; therefore, the width
of $\mbox{CI}_0$ shrinks to zero as does coverage since there is
remaining metamodel uncertainty in general. However, because
$\mbox{CI}_+$ accounts for metamodel uncertainty it still provides
asymptotically consistent coverage.  This effect is demonstrated by
the empirical study in Section~\ref{sec:empirical}.

Our decomposition allows us to express the total variance in
Equation~(\ref{eq.varDecomp}) as the sum of two variances measuring
input and metamodel uncertainty: $\sigma^2_T = \sigma^2_I+\sigma^2_M.$
In the metamodel-assisted bootstrapping framework, we can estimate
each variance component as follows:
\begin{itemize}

\item Total variance:
$\widehat{\sigma}^2_T=\sum_{b=1}^B(M_b-\bar{M})^2/(B-1)$,
where $\bar{M}=\sum_{b=1}^B M_b/B.$

\item Input variance:
$\widehat{\sigma}^2_I=\sum_{b=1}^B(\textcolor{black}{\mu}_b-\bar{{\mu}})^2/(B-1)$,
where $\bar{{\mu}}=\sum_{b=1}^B\textcolor{black}{\mu}_b/B$.  

\item Metamodel variance:
$\widehat{\sigma}^2_M= \sum_{b=1}^B\textcolor{black}{\sigma}_p^2
(\widehat{\mathbf{X}}^{(b)}_{\mathbf{m}})/B$.
\end{itemize}

The ratio $\widehat{\sigma}_I/\widehat{\sigma}_T$ provides an estimate
of the relative contribution from input uncertainty on $\mbox{CI}_+$.
If it is close to 1, the contribution from metamodel uncertainty can
be ignored.  Thus, this ratio can help a decision maker determine
where to put more effort: If the input variance dominates, then get
more real-world data (if possible). If the metamodel variance
dominates, then it can be reduced by more simulation, which can be a
combination of additional design points and additional replications at
existing design points. If neither dominates, then both activities are
necessary to reduce $\mbox{CI}_+$ to a practically useful size.

The asymptotic properties of these variance component estimators
 are shown in the following theorems.

\begin{theorem} \label{thm:decomposition1}
Suppose that Assumptions~1--4 hold. Then
conditional on $\bar{\mathbf{Y}}_\mathcal{D}$ the variance component
estimators $\widehat{\sigma}^2_M, \widehat{\sigma}^2_I,
\widehat{\sigma}^2_T$ are consistent as $m, B\rightarrow\infty$, where
as $m \rightarrow \infty$ we have $m_\ell/m \rightarrow c_\ell$,
$\ell=1,2,\ldots,L$, for a constant $c_\ell>0$. Specifically, 
\begin{itemize}

\item As $m\rightarrow\infty$, the input uncertainty disappears:
\[
\lim_{m\rightarrow\infty}\sigma^2_M=\sigma^2_{p}(\bx_c),
\lim_{m\rightarrow\infty}\sigma^2_I=0 \mbox{ and
}\lim_{m\rightarrow\infty}\sigma^2_T=\sigma^2_{p}(\bx_c).
\]

\item As $m\rightarrow\infty$ and $B\rightarrow\infty$ in an iterated
limit, the variance component estimators are consistent: 
\begin{eqnarray*}
\lim_{m\rightarrow\infty}\lim_{B\rightarrow\infty}\widehat{\sigma}_M^2
&=&\lim_{m\rightarrow\infty}\sigma_M^2=\sigma_p^2(\bx_c), \\
\lim_{m\rightarrow\infty}\lim_{B\rightarrow\infty}\widehat{\sigma}_I^2
&=&\lim_{m\rightarrow\infty}\sigma_I^2=0, \\
\lim_{m\rightarrow\infty}\lim_{B\rightarrow\infty}\widehat{\sigma}_T^2
&=&\lim_{m\rightarrow\infty}\sigma_T^2=\sigma^2_p(\bx_c). 
\end{eqnarray*}

\end{itemize}

\end{theorem}

Theorem~\ref{thm:decomposition1} demonstrates that the variance
components estimators $\widehat{\sigma}^2_I$, $\widehat{\sigma}^2_M$
and $\widehat{\sigma}^2_T$ are consistent. However, we can see that
when $m\rightarrow\infty$ the input uncertainty disappears. Since
$\lim_{m\rightarrow\infty}\sigma^2_I=\lim_{m\rightarrow\infty}
\lim_{B\rightarrow\infty}\widehat{\sigma}^2_I=0$ is not interesting,
we study the consistency of scaled versions of $\sigma^2_I$ and
$\widehat{\sigma}^2_I$ in Theorem~3, showing that $m\sigma^2_I$ and
$m\widehat{\sigma}_I^2$ converge to the same non-zero constant.

\begin{theorem} \label{thm:decomposition2}
Suppose that Assumptions~1--4 and the
following additional assumptions hold:
\setcounter{enumi}{4}
\begin{enumerate}

\item [5.] The first three derivatives of the correlation function of the
GP $M(\bx)$ exist and the third derivative is bounded; and

\item [6.] $m_\ell/m \rightarrow 1$ for $\ell=1,2,\ldots,L$.

\end{enumerate}
Then $\lim_{m\rightarrow\infty}m\sigma_I^2=\lim_{m\rightarrow\infty}
\lim_{B\rightarrow\infty} m\widehat{\sigma}^2_I=\sigma^2_\mu$ almost
surely, where $\sigma^2_\mu$ is a positive constant.

\end{theorem}

Theorems~\ref{thm:decomposition1}--\ref{thm:decomposition2} give the
asymptotic properties of the variance component estimators,
guaranteeing $\widehat{\sigma}_I/\widehat{\sigma}_T$ is a consistent
estimator for the relative contribution of input to overall
uncertainty.  We will empirically evaluate its finite-sample
performance in Section~\ref{sec:empirical} \textcolor{black}{where we
form the variance component estimators by inserting
$(\widehat{\tau}^2, \widehat{\pmb{\theta}}, \widehat{C})$ for the
unknown parameters $(\tau^2, \pmb{\theta}, C)$.}

\subsection{Unstable and Undefined Moments}

A fundamental assumption of simulation is that the expectation
$\mu(\bx_c)$ exists. This assumption does not imply, however, that it
exists for \textit{all} possible values of $\bx$, $\bX_\bm$ or
$\widehat{\bX}_\bm^{(b)}$ that might be realized. The prototype
example is a congestion-related performance measure of a queueing
system as time goes to infinity when congestion increases without
bound for some values of its interarrival-time and service-time
parameters. We refer to systems for which $\mu(\bx)$ is $\pm \infty$
for some values of $\bx$ as potentially \textit{unstable}.

Recall that $\mu(\bx) = \E[Y(\bx)]$. A second problem arises when for
some values of $\bx$ the random variable $Y(\bx)$ is undefined.
The prototype example is a network for which we want to estimate some
start-to-finish performance measure, but the start and finish are
not connected for certain values of $\bx$. We refer to such systems as
potentially \textit{undefined}.

Below we use illustrative examples to describe what happens to
metamodel-assisted bootstrapping in each case, why we expect to be
robust to unstable systems, and what needs to be done for undefined
systems.  We assume that $\bx_c$ is an interior point of the space
\textcolor{black}{$\Psi$ for
which $\mu(\bx_c)$ is stable and $Y(\bx_c)$ is defined} so both problems
disappear asymptotically ($m \rightarrow \infty$), but they may occur
when we apply the metamodel-assisted bootstrapping approach to a
finite sample of real-world data.

\subsubsection{Unstable Moments}
\label{sec:unstable}

Consider the simulation of an $M/M/1$ queue. Let $x_1$ and $x_2$
denote the mean interarrival time and mean service time, respectively,
and let $\bx = (x_1, x_2)^\top$. The unknown mean response $\mu(\bx)$
is the steady-state expected number of customers in the system. The
true values $x_1^c$ and $x_2^c$ are unknown and must be estimated from
real-world data; $x_1^c > x_2^c$ so the system is actually stable.  We
denote the unstable and stable regions of $\bx$ by $U = \{(x_1, x_2):
0 < x_1 \le x_2 \}$ and $\bar{U} = \{(x_1, x_2): x_1 > x_2 > 0\}$,
respectively, and $\bx_c$ is an interior point of $\bar{U}$.

% so that $\mu(\bx)<\infty$ if $\bx\in \bar{U}$ and $\mu(\bx)=\infty$ if
%$\bx\in U$.

%Although the system is stable at $\bx_c$, it may not be stable for
%every possible design point or bootstrap moment.  

As described in the Appendix, we use an initial set of bootstrap
resampled moments to define an ellipsoid in which to embed our
experiment design to fit the metamodel, and then generate a second set
at which we evaluate the metamodel to form a two-sided,
equal-probability bootstrap percentile interval.  The conditional
probability that a bootstrap resampled moment
$\widehat{\mathbf{X}}_{\mathbf{m}}^{(b)}$ is located in the unstable
region given the real-world data is
\begin{equation}
 P_U \equiv \mbox{Pr}\left\{\left.
\widehat{\mathbf{X}}_{\mathbf{m}}^{(b)} \in U \right| \mathbf{z}_{\mathbf{m}}^{(0)}
\right\}.
\label{eq.infP}
\end{equation}

%Since we quantify input uncertainty using a two-sided,
%equal-probability bootstrap percentile CI, the value of $P_U$ impacts
%the CI. If $P_U\geq\alpha/2$, then the $\widehat{\bX}_\bm^{(b)}$ value
%that corresponds to the upper confidence bound is likely in $U$;
%otherwise, the $\widehat{\bX}_\bm^{(b)}$ values corresponding to both
%the lower and upper confidence bounds are likely in the stable region.

For the $M/M/1$ queue we know $U$ so we know which bootstrap moments
are in the unstable region; therefore, we could map the mean response
for unstable moments (symbolically) to $\infty$ and only fit or use
the metamodel to predict the mean response at stable moments. If $P_U$
is large this could lead to a one-sided lower confidence interval
(infinite upper limit) that would be interpreted as ``the real system
may be unstable;''  this is an appropriate
conclusion if input uncertainty is substantial.

Unfortunately, in general stochastic systems it could be difficult or
impossible to determine which moments are in the unstable region
either analytically or empirically
\citep{Wieland_Pasupathy_Schmeiser_2003}. Thus, in the experiment
design phase we might simulate the system at an $\bx \in U$ to fit the
metamodel, and in the bootstrapping phase we might evaluate the
resulting metamodel at an $\bx \in U$ to estimate the CI. What is the
effect of doing this when $P_U>0$, possibly even large?

Suppose we start each replication of the $M/M/1$ queue with an empty
and idle system. Let $\mu(\bx,t)$ denote the true expected number of
customers in the system at time $t \ge 0$. Then except for the case
$x_1 = x_2$, which we ignore, it can be shown that $\mu(\bx,t)$
satisfies the differential equation
\[ 
\frac{d\mu(\bx,t)}{dt}=\frac{1}{x_1}-\frac{1-p_0(t)}{x_2} 
\]
where $p_0(t)$ is the probability that the system is empty at $t$
and $p_0(0)=1$. If $\bx\in \bar{U}$ then $p_0(t)\rightarrow 1-x_2/x_1$ as
$t\rightarrow \infty$; however, if $\bx\in U$ then $p_0(t)\rightarrow
0$ as $t\rightarrow\infty$. Thus, for large $t$, 
\begin{equation}
\frac{d\mu(\bx,t)}{dt} \approx \left\{\begin{array}{ll}
0, & \mbox{if $\bx\in \bar{U}$}  \\
\label{eq.unstable}
\frac{1}{x_1}-\frac{1}{x_2} > 0, & \mbox{if $\bx\in U$}.
\end{array}
\right. 
\end{equation}
For any finite run length $T$ and warm-up period $T_0 < T$ the simulation
provides an unbiased estimator of
\begin{equation}
\label{eq:transient.mean}
\bar{\mu}(\bx, T_0, T) = \frac{1}{T - T_0}\int_{T_0}^T \mu(\bx, t)\,dt .
\end{equation}
Notice that this quantity is finite for any positive values of $x_1$
and $x_2$, whether stable or not. However, if $\bx \in \bar{U}$ then
$\bar{\mu}(\bx, T_0, T)$ converges to $\mu(\bx)$ for large $T$; while
if $\bx \in U$ then $\bar{\mu}(\bx, T_0, T)$ is increasing in $T$ for
$T$ large enough.  

\textit{The key point is this: The expected value of any
simulation-based estimator will be finite, even if the true
steady-state mean is not. Further, the expected value of the
simulation estimator at unstable $\bx$ will tend to be larger than at
near-by stable $\bx$. This means that the simulation estimates
corresponding to unstable $\bx$ will tend to be the largest ones
observed, but still not infinite.}

Consider the design points or bootstrap resampled moments that are in
$U$. When $P_U < \alpha/2$ and the run length is long enough, the
unstable design points used to fit the metamodel, or bootstrap moments
at which it is evaluated, tend not to adversely affect either the
metamodel or the interval estimate because they are in the right tail
beyond the $\alpha/2$ quantile. On the other hand, when $P_U \ge
\alpha/2$ the large estimates corresponding to unstable design points
or bootstrap moments tend to lengthen the interval estimate beyond
what is required to cover $\mu(\bx_c)$; this causes overcoverage
rather than undercoverage.  Thus, metamodel-assisted bootstrapping
\textcolor{black}{will often be} robust to unstable moments
\textcolor{black}{in the sense of not being too short, but possibly too
long; we demonstrate this empirically in Section~6.}

\subsubsection{Undefined Moments}
\label{sec:infeasible}

Consider the queueing network example in Figure~\ref{fig:Figure1_3}.
For simplicity, suppose that the routing probabilities
$p_1,p_2,p_3$ are the only input parameters. Let
$\bx=(x_1,x_2,x_3)^\top = (p_1,p_2,p_3)^\top$.  The true parameters
$\bx_c$ are unknown and estimated by finite samples from Bernoulli
distributions ($1$ if the customer takes a particular route, $0$
otherwise). Suppose that the mean response of interest, $\mu(\bx),$ is
the steady-state expected time for a customer to traverse the network,
which exists and is well-defined at $\bx_c$.  Unfortunately, $Y(\bx)$ may
not be defined for every possible bootstrap resampled moment $\bx$.  For
instance, if $\widehat{\bX}_{\bm}^{(b)} = (0, 0.665, 0)^\top$ then
Stations~1 (start) and~4 (end) are disconnected and no simulation
output for time to traverse the network will ever be generated.  Thus,
the system corresponding to this bootstrap moment is undefined. 

In practical problems for which we can obtain real-world input data,
we should know a priori that the system performance measure is well
defined (e.g., we would not include a route unless we actually
observed a customer take it). Further, it should not be difficult to
detect moments for which the system output is undefined, either
because we understand the system logic (as in this example) or because
the simulation simply fails to run. Therefore, a reasonable solution
to the problem of undefined systems is to reject (and sample again)
bootstrap moments $\widehat{\bX}_{\bm}^{(b)}$ that imply an undefined
output. This makes our assessment of uncertainty \textit{conditional}
on the system performance measure being defined, which makes
sense.

\section{Empirical Study}
\label{sec:empirical}

In this section we use the queueing network described in
Section~\ref{sec:problem_Description} to evaluate the performance of
our metamodel-assisted bootstrapping approach. The performance measure
is the steady-state expected number of customers in the system.  Both
interarrival and service times follow gamma distributions and the
routing decisions follow Bernoulli distributions. Thus, it is a
13-dimensional problem with $L=8$ input processes that include both
continuous and discrete distributions. The true parameters of the
input distributions are $\alpha_A=1,\beta_A=0.25$, $\alpha_{S_i}=1$,
$\beta_{S_i}=0.2$ for $i=1,2,3,4$ and $p_1=p_2=0.5, p_3=0.75$. These
parameter values imply a tractable Jackson network with steady-state
number of customers in system $\mu(\bx_c) = 12.67$. The maximum
traffic intensity at any station is $0.8$.

In the experiments we assume that all parameters for all input
distributions are unknown and are estimated from a finite sample of
real-world data. Notice that $\alpha_A,\beta_A$, $\alpha_{S_i}$,
$\beta_{S_i}$ for $i=1,2,3,4$ are estimated from continuous
measurements, while the routing probabilities $p_1,p_2,p_3$ are
estimated from 0 or 1 observations that would correspond to customer
routing decisions.  The model with \textit{estimated} input parameters
is almost surely not a Jackson network and it could be unstable. Our
measure of uncertainty is a $95\%$ CI for $\mu(\bx_c)$ \textcolor{black}{as defined
by~(\ref{eq.CI_2}) because this is the objective desired in practice.}

\begin{figure*}[tb]
\vspace{0.2in}
{
\centering
\includegraphics[scale=0.35]{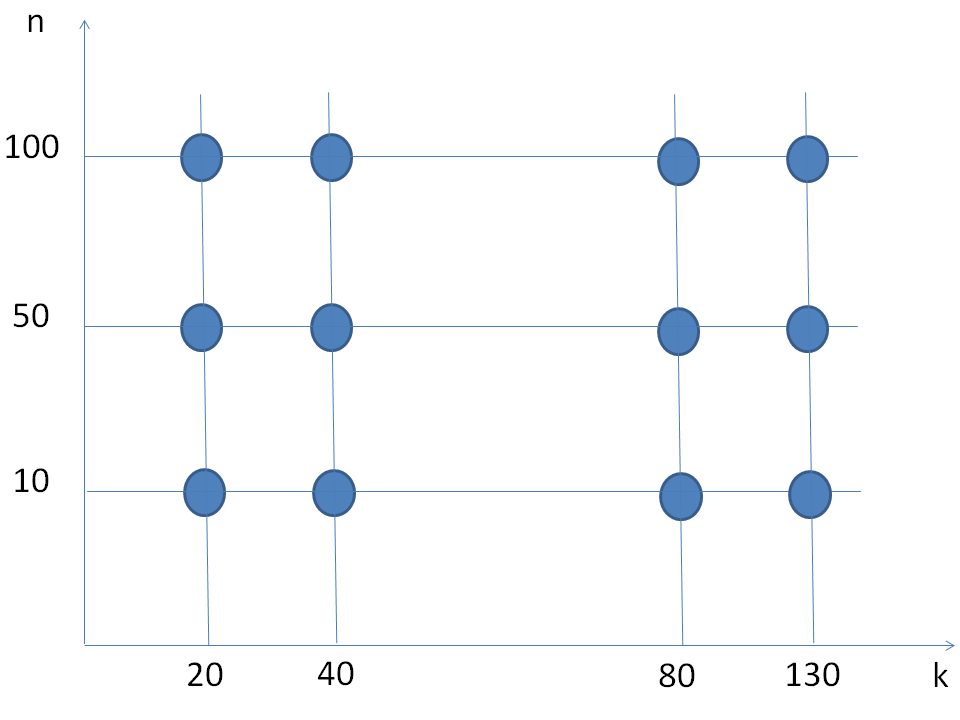}
\vspace{-0.2in}
\caption{Experiment design.
\label{fig:Figure2_2}}
}
\end{figure*}

To evaluate the robustness of the metamodel-assisted bootstrapping
approach, we systematically examine the effect of the quantity of
real-world data and the number of design points and replications per
design point used to fit the metamodel; see
Figure~\ref{fig:Figure2_2}. We consider a wide range for the quantity
of real-world data $m=50,500,5000$, letting $m_\ell=m$ for
$\ell=1,2,\ldots L$. The levels for the number of design points are
$k=20,40,80,130$. For a 13-dimensional problem $k=20$ is a very small
design. The studies by \cite{Jones_1998} and \cite{Loeppky_2009}
recommend that the number of design points should be $10$ times the
dimension of the problem for kriging; we take this as the maximum
number of design points. The same number of replications are assigned
to all design points and we try $n=10,50,100$. 

\cite{barton_nelson_xie_2011} demonstrated that $\mbox{CI}_0$ has good
performance when the impact of metamodel uncertainty is negligible.
In this empirical study we focus on situations where metamodel
uncertainty may be significant. However, rather than creating a
problem that actually takes hours or days to run, we instead construct
a problem with high metamodel uncertainty by using short run lengths
for each replication: 20 time units after the warm up, which is
roughly equivalent to 80 finished customers. To avoid the influence
from initial bias, all simulations start loaded with the number of
customers at each station being their steady-state expected values
(rounded) under $\bx_c$.  Furthermore, a long warmup period of 200
time units is used. The net effect is that the point estimators of the
steady-state number in the network have low bias, but may be quite
variable.

To make the description of the empirical results easy to follow, we
start with overall conclusions:
\begin{enumerate}

\item The new ACI $\mbox{CI}_+$ is robust to different
levels of real-world data $m$, number of design points $k$ and number
of replications $n$.  

\item When metamodel uncertainty is significant, $\mbox{CI}_0$ tends
to have undercoverage that becomes more serious as $m$ increases.
Since $\mbox{CI}_+$ accounts for  metamodel uncertainty, it does
not exhibit this degradation although it sometimes has slight
overcoverage.  

\item Metamodel-assisted bootstrapping \textcolor{black}{continues to
deliver at least the nominal coverage} when the probability of an
unstable system $P_U$ is large.  

\item The ratio $\widehat{\sigma}_I/\widehat{\sigma}_T$ is a useful
measure of the relative contribution of input to overall statistical
uncertainty.

\end{enumerate}

As discussed in Section~\ref{sec:unstable}, metamodel-assisted
bootstrapping might behave differently when $P_U < \alpha/2$ vs.\ $P_U
\ge \alpha/2$. Since $P_U$ only depends on $m$ and $\bx_c$, we ran a
side experiment to estimate it using
\begin{equation}
\label{eq.infP_hat}
\widehat{P}_U=\frac{1}{B}\sum_{b=1}^B\mbox{I}
\left(\widehat{\mathbf{X}}_{\mathbf{m}}^{(b)} \in U\right),
\end{equation}
where $\mbox{I}(\cdot)$ is the indicator function. The means and
standard deviations (SD) of $\widehat{P}_U$ for $m=50,500,5000$ were
estimated based on 1000 macro-replications and are displayed in
Table~\ref{table:infP}. In each macro-replication we independently
generated a sample of size $m$ of ``real-world data.''  Then,
conditional on these data, we drew $B=2000$ bootstrap resampled
moments. Finally, we calculated the estimate of $\widehat{P}_U$ using
Equation~(\ref{eq.infP_hat}). 

As $m$ increases the bootstrap resampled moments become more closely
centered around $\bx_c$.  Thus, both the mean and SD of
$\widehat{P}_U$ decrease with increasing $m$ as shown in
Table~\ref{table:infP}. When $m=50$, ${P}_U$ appears to be much larger
than $\alpha/2$ so the bootstrap moments
$\widehat{\mathbf{X}}_{\mathbf{m}}^{(b)}$ that correspond to the upper
confidence bound are located in the unstable region $U$ with high
probability. When $m=500$, ${P}_U$ appears to be close to
$\alpha/2=2.5\%$, while when $m=5000$ there is little chance of
getting unstable bootstrap moments.

\begin{table}[tb]
\caption{Percentage of unstable bootstrap resampled moments.}
\label{table:infP}
\begin{center}
\begin{tabular}{|c|c|c|c|} 
\hline
& $m=50$ & $m=500$ & $m=5000$  \\
 \hline
mean of $\widehat{P}_U$  & 44.4\% & 2.3\% & 0 \\ \hline
SD of $\widehat{P}_U$  & 31.7\% & 7.9\% & 0 
\\ \hline
\end{tabular}
\end{center}
\end{table}

In the following sections we describe the overall performance of
$\mbox{CI}_0$ and $\mbox{CI}_+$, \textcolor{black}{including the situation where 
$P_U>0$}, and analyze the finite-sample performance of
$\widehat{\sigma}_I/\widehat{\sigma}_T$ as a measure of the relative
contribution of input to overall uncertainty.

\subsection{Performance of CIs}
\label{subSec:empirical_CI}

Tables~\ref{table:HDm50}--\ref{table:HDm500-5000} show the results for
$\mbox{CI}_0$ and $\mbox{CI}_+$ when $m=50,500,5000$, including the
probability of covering $\mu(\bx_c)$, and the mean and SD of the interval
widths. All results are based on $1000$ macro-replications. 

When $m=50$, $P_U$ is much greater than $\alpha/2$ according to
Table~\ref{table:infP}. This explains the very large CI widths in
Table~\ref{table:HDm50}. Nevertheless, both $\mbox{CI}_0$ and
$\mbox{CI}_+$ have reasonable coverage overall, an observation we
explore further in Section~\ref{subSec:empirical_Unstable}.  Notice
that $\mbox{CI}_0$ does exhibit undercoverage when we use a very small
experiment design of $k=20$ points, while the coverage of
$\mbox{CI}_+$ is much closer to the nominal value of $95\%$ in this
case. If we fix the number of replications $n$ and increase the number
of design points $k$, the coverage of $\mbox{CI}_0$ improves. For a
fixed $k$ the effect of increasing $n$ is not as obvious.

Table~\ref{table:HDm500-5000} shows the results for $m=500,5000$.
Compared with the results for $m=50$, the mean and SD of the interval
widths drop dramatically.  The effects of $k$ and $n$ are easier to
discern especially when $m=5000$, which has no unstable bootstrap
moments.  Specifically, for a fixed quantity of real-world data $m$,
if either the number of design points $k$ or replications per design
point $n$ is small then $\mbox{CI}_0$ tends to have undercoverage
because it fails to account for substantial metamodel uncertainty,
unlike $\mbox{CI}_+$. However, because $\mbox{CI}_+$ does incorporate
metamodel uncertainty it sometimes has slight overcoverage.  

The most troubling observation about $\mbox{CI}_0$ is that, for fixed
$(n,k)$, as the amount of input data $m$ increases its undercoverage
becomes more serious.  The diminished coverage occurs because as
$m\rightarrow\infty$ the width of $\mbox{CI}_0$ shrinks to zero, which
is not appropriate when there is still metamodel uncertainty. Again,
$\mbox{CI}_+$ does not exhibit this degradation. As $n$ and $k$
increase, the coverages of $\mbox{CI}_0$ and $\mbox{CI}_+$ become
closer to each other.

\textit{The behavior of $\mbox{CI}_0$ is what we would expect based on
\cite{barton_nelson_xie_2011}, which introduced $\mbox{CI}_0$. Their
procedure continued to add simulation effort (design points and
replications) until its effect on the confidence interval was
negligible. \textcolor{black}{Compared to $\mbox{CI}_0$, the new
interval, $\mbox{CI}_+$, is able to account for the effect of the
remaining simulation estimation error. Therefore, it can work under
more general situations where the simulated systems are complex and
the simulation budget is tight.}}

\begin{table}[tb]
\caption{Results for $\mbox{CI}_{0}$, $\mbox{CI}_{+}$ and $\widehat{\sigma}_I/\widehat{\sigma}_T$
when $m=50$. } \label{table:HDm50}
\begin{center}
\begin{tabular}{|c|c|c|c|c|c|c|} 
\hline
$m=50$ & \multicolumn{3}{c|}{$k=20$} & \multicolumn{3}{c|}{$k=40$} \\ \cline{2-7}
 &  $n=10$ & $n=50$  & $n=100$ & $n=10$ & $n=50$  & $n=100$  \\
 \hline
Coverage of $\mbox{CI}_0$ & 91.9\% & 92.3\% & 91.5\% & 93.8\%  & 94.4\% & 93.4\%\\ \hline
Coverage of $\mbox{CI}_{+}$ & 93.9\% & 94.9\% & 93.7\% & 94.9\% & 95.6\% & 95.9\% \\ \hline
$\mbox{CI}_0$ Width (mean)& 326.4 & 332.4 & 339.5 & 319.1 & 328.6 & 326.5 
\\ \hline 
$\mbox{CI}_{+}$ Width (mean) & 344.1 & 348.8 & 357.1 & 332.3 & 342.3 & 341.2 
\\ \hline 
$\mbox{CI}_0$ Width (SD) & 183.1 & 173.6 & 180.7 & 176.4 & 167.6 & 175  
\\ \hline 
$\mbox{CI}_{+}$ Width (SD) & 188 & 175.7 & 183.8 & 178.2 & 169.2 & 176.1   
\\ \hline 
$\widehat{\sigma}_I/\widehat{\sigma}_T$ & 0.963 & 0.965 & 0.964 & 0.973 & 0.973 & 0.971
\\ \hline\hline
$m=50$ & \multicolumn{3}{c|}{$k=80$} & \multicolumn{3}{c|}{$k=130$} \\ \cline{2-7}
 &  $n=10$ & $n=50$  & $n=100$ & $n=10$ & $n=50$  & $n=100$  \\
 \hline
Coverage of $\mbox{CI}_0$ & 94.6\% & 96.3\% & 95.4\% & 94.2\%  & 95.1\% & 95.4\%\\ \hline
Coverage of $\mbox{CI}_{+}$ & 95.9\% & 96.7\% & 96.1\% & 94.5\% & 96\% & 96.1\% \\ \hline
$\mbox{CI}_0$ Width (mean) & 312.1 & 314.8 & 322.7 & 322 & 321.86 & 320
\\ \hline  
$\mbox{CI}_{+}$ Width (mean) & 322 & 325.7 & 334 & 330.2 & 331 & 329.4
\\ \hline 
 $\mbox{CI}_0$ Width (SD) & 169.7 & 159.1 & 164.7 & 171.5 & 169.3 & 172.3  
\\ \hline 
$\mbox{CI}_{+}$ Width (SD) & 171.2 & 159.4 & 165 & 172.7 & 169.5 & 172.7 
\\ \hline
$\widehat{\sigma}_I/\widehat{\sigma}_T$ & 0.982 & 0.98 & 0.978 & 0.985 & 0.985 & 0.983
\\ \hline
\end{tabular}
\end{center}
\end{table}

\begin{table}[tb]
\caption{Results for $\mbox{CI}_{0}$, $\mbox{CI}_{+}$ and $\widehat{\sigma}_I/\widehat{\sigma}_T$
when $m=500$ and $m=5000$. }
\label{table:HDm500-5000}
\begin{center}
\begin{tabular}{|c|c|c|c|c|c|c|} 
\hline
$m=500$ & \multicolumn{3}{c|}{$k=20$} & \multicolumn{3}{c|}{$k=40$} \\ \cline{2-7}
 &  $n=10$ & $n=50$  & $n=100$ & $n=10$ & $n=50$  & $n=100$  \\
 \hline
Coverage of $\mbox{CI}_0$ & 90.5\% & 94.6\% & 95.1\% & 94.9\%  & 96.7\% & 96.4\%\\ \hline
Coverage of $\mbox{CI}_{+}$ & 95.7\% & 97.7\% & 97.8\% & 96.6\% & 98.3\% & 97.8\% \\ \hline
$\mbox{CI}_0$ Width (mean)& 24.8 & 28.1 & 29.4 & 27.1 & 28.5 & 28.7
\\ \hline 
$\mbox{CI}_{+}$ Width (mean) & 28.9 & 30.8 & 32.2 & 29.6 & 30.3 & 30.5
\\ \hline
$\mbox{CI}_0$ Width (SD) & 19.9 & 19.4 & 20.6 & 19.1 & 19.2 & 19.9 \\\hline
$\mbox{CI}_+$ Width (SD) & 20.6 & 20.4 & 21.7 & 19.7 & 19.9 & 20.6 \\ \hline 
$\widehat{\sigma}_I/\widehat{\sigma}_T$ & 0.88 & 0.932 & 0.933 & 0.932 & 0.957 & 0.958
\\ \hline\hline
$m=500$ & \multicolumn{3}{c|}{$k=80$} & \multicolumn{3}{c|}{$k=130$} \\ \cline{2-7}
 &  $n=10$ & $n=50$  & $n=100$ & $n=10$ & $n=50$  & $n=100$  \\
 \hline
Coverage of $\mbox{CI}_0$ & 96.5\% & 97.5\% & 95.8\% & 95.4\%  & 96.5\% & 95.9\%\\ \hline
Coverage of $\mbox{CI}_{+}$ & 98\% & 98.3\% & 97.3\% & 97.5\% & 97.1\% & 96.9\% \\ \hline
$\mbox{CI}_0$ Width (mean) & 26.3 & 28 & 28.7 & 26.4 & 27.9 & 27.6
\\ \hline 
$\mbox{CI}_{+}$ Width (mean) & 28 & 29 & 29.7 & 27.9 & 28.6 & 28.2
\\ \hline
$\mbox{CI}_0$ Width (SD) & 17.4 & 18 & 19.3 & 18.8 & 19.6 & 19.3 \\ \hline
$\mbox{CI}_+$ Width (SD) &  17.7 & 18.4 & 19.6 & 18.9 & 19.9 & 19.5 \\ \hline
$\widehat{\sigma}_I/\widehat{\sigma}_T$ & 0.952 & 0.977 & 0.978 & 0.957 &0.984 & 0.987
\\ \hline\hline\hline
 $m=5000$ & \multicolumn{3}{c|}{$k=20$} & \multicolumn{3}{c|}{$k=40$} \\ \cline{2-7}
 &  $n=10$ & $n=50$  & $n=100$ & $n=10$ & $n=50$  & $n=100$  \\
 \hline
Coverage of $\mbox{CI}_0$ & 70.7\% & 89.2\% & 93.1\% & 81.5\%  & 94.3\% & 94.8\%\\ \hline
Coverage of $\mbox{CI}_{+}$ & 91.3\% & 96.3\% & 95.6\% & 96.5\% & 96.1\% & 96.3\% \\ \hline
$\mbox{CI}_0$ Width (mean)& 3.29 & 3.97 & 4.14 & 3.93 & 4.23 & 4.3
\\ \hline 
$\mbox{CI}_{+}$ Width (mean) & 5.85 & 4.8 & 4.56 & 6.08 & 4.64 & 4.52
\\ \hline 
$\mbox{CI}_0$ Width (SD) & 1.89 & 1.2 & 1 & 1.64 & 0.87 & 0.83 \\ \hline
$\mbox{CI}_+$ Width (SD) & 2.12 & 1.13 & 1 & 1.52 & 0.89 & 0.85 \\ \hline
$\widehat{\sigma}_I/\widehat{\sigma}_T$ & 0.588 & 0.85 & 0.924 & 0.664 & 0.924 & 0.959
\\ \hline\hline
$m=5000$ & \multicolumn{3}{c|}{$k=80$} & \multicolumn{3}{c|}{$k=130$} \\ \cline{2-7}
 &  $n=10$ & $n=50$  & $n=100$ & $n=10$ & $n=50$  & $n=100$  \\
 \hline
Coverage of $\mbox{CI}_0$ & 88.9\% & 93.6\% & 94.9\% & 89.5\%  & 93.7\% & 94.8\%\\ \hline
Coverage of $\mbox{CI}_{+}$ & 98.1\% & 95\% & 96\% & 98\% & 95.6\% & 95.5\% \\ \hline
$\mbox{CI}_0$ Width (mean) & 4.54 & 4.29 & 4.29 & 4.52 & 4.35 & 4.32
\\ \hline 
$\mbox{CI}_{+}$ Width (mean) & 6.1 & 4.56 & 4.42 & 5.98 & 4.64 & 4.45
\\ \hline
$\mbox{CI}_0$ Width (SD) & 1.37 & 0.85 & 0.77 & 1.28 & 0.9 & 0.79 \\ \hline
$\mbox{CI}_+$ Width (SD) & 1.27 & 0.85 & 0.78 & 1.13 & 0.87 & 0.77 \\ \hline
$\widehat{\sigma}_I/\widehat{\sigma}_T$ & 0.757 & 0.946 & 0.974 & 0.766 & 0.945 & 0.974
\\ \hline
\end{tabular}
\end{center}
\end{table}

\subsection{Performance of $\widehat{\sigma}_I/\widehat{\sigma}_T$}
\label{subSec:empirical_Ratio}

\begin{figure*}[tb]
\vspace{0.2in}
{
\centering
\includegraphics[scale=0.7]{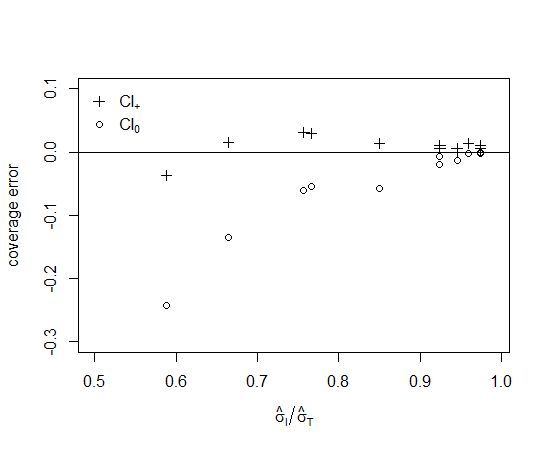}
\vspace{-0.2in}
\caption{The coverage errors for $\mbox{CI}_0$ and $\mbox{CI}_+$ 
vs.\ $\widehat{\sigma}_I/\widehat{\sigma}_T$ when $m=5000$ across
all values of $n$ and $k$.
\label{fig:coverageErr_ratio_m5000}}
}
\end{figure*}

Tables~\ref{table:HDm50}--\ref{table:HDm500-5000} also demonstrate
that $\widehat{\sigma}_I/\widehat{\sigma}_T$ provides a good measure
of the relative contribution of input to overall uncertainty, and
behaves as it should:
\begin{itemize}

\item For a fixed amount of real-world data $m$, increasing the number
of design points and replications $(n,k)$ drives
$\widehat{\sigma}_I/\widehat{\sigma}_T$ toward 1, indicating a
decrease in metamodel uncertainty.

\item For fixed simulation effort $(n,k)$, increasing the amount of
real-world data $m$ decreases $\widehat{\sigma}_I/\widehat{\sigma}_T$,
indicating that there is relatively less input uncertainty. Notice,
however, that the relationship is not simple because as $m$ increases
the design space over which we fit the metamodel becomes smaller, so
that even with the same simulation effort the absolute level of
metamodel uncertainty will decrease somewhat. 

\item When $\widehat{\sigma}_I/\widehat{\sigma}_T$ is near $1$, the
behaviors (coverage and width) of $\mbox{CI}_0$ and $\mbox{CI}_{+}$
are similar and both have coverage close to the nominal level; this is
illustrated in Figure~\ref{fig:coverageErr_ratio_m5000}. Recall that
$\mbox{CI}_0$ does not account for metamodel uncertainty, and that
$\widehat{\sigma}_I/\widehat{\sigma}_T \approx 1$ indicates that input
uncertainty is large relative to metamodel uncertainty, which is when
$\mbox{CI}_0$ will do best. Figure~\ref{fig:coverageErr_ratio_m5000}
also illustrates the general robustness of $\mbox{CI}_+$.

\end{itemize}

\subsection{Robustness to Unstable Moments}
\label{subSec:empirical_Unstable}

Recall the observation from Table~\ref{table:HDm50} that when there is
a small quantity of real-world data ($m=50$), resulting in a large
probability of unstable bootstrap moments, then both $\mbox{CI}_0$ and
$\mbox{CI}_+$ had large mean and SD of their widths, yet provided
reasonable coverage. Examining the results, we found that most of the
intervals that fail to cover the mean do so because the lower
confidence bound is \textit{above} $\mu(\bx_c)$; this is the case for
both $\mbox{CI}_0$ and $\mbox{CI}_+$.  Using all 1000
macro-replications, the estimated probability that the lower
confidence bound is above the mean (greater than $\mu(\bx_c)$, meaning
too large) is 4.4\% for $\mbox{CI}_0$ and 3.8\% for $\mbox{CI}_+$,
while the estimated probability that the upper confidence bound is below
the mean (less than $\mu(\bx_c)$, meaning too small) is only $0.2\%$
for $\mbox{CI}_0$ and $0.1\%$ for $\mbox{CI}_+$; for two-sided equal percentile
intervals we would expect these to be around $2.5\%$.  

%The simulation outputs for unstable systems have very large mean and
%variance, so the mean and variance of CI width is correspondingly
%large when the percentage of unstable bootstrap moments is
%substantial. 

\textcolor{black}{\textit{We conclude that even though the metamodel
predicts a finite mean when it should be infinite, this will still
tend to lead to overcoverage rather than undercoverage, and therefore
is conservative.}}

\section{Conclusions}

In this paper, a metamodel-assisted bootstrapping approach is used for
statistical uncertainty analysis.  Input uncertainty is approximated
by the bootstrap, an equation-based stochastic kriging metamodel is
used to propagate the input uncertainty to the output mean, and the
metamodel uncertainty is derived using properties of stochastic
kriging.  This approach delivers an interval estimator that accounts
for \textit{all} statistical uncertainty, both simulation and input.
The asymptotic consistency of this interval is proved under the
assumption that the true response surface is a realization of a
Gaussian process and certain parameters are known.

An empirical study on a difficult problem demonstrates that our
approach \textcolor{black}{can have} good finite-sample performance even
when there are several input distributions (both discrete-valued and
continuous-valued), a tight computational budget, and bootstrap
moments corresponding to unstable systems. Thus, the new interval does not
require a sequential experiment to make metamodel uncertainty
negligible, as in \cite{barton_nelson_xie_2011}.

If $\mbox{CI}_+$ is too wide, then it is important to know the
relative contributions from input and metamodel uncertainty as a guide
to either collecting more real-world data or doing more simulation or
both. We give a measure of the relative contribution of input to
overall statistical uncertainty by using a variance decomposition and
analyze its asymptotic properties.

\section*{Acknowledgments}

This paper is based upon work supported by the National Science
Foundation under Grant Nos.\ CMMI-0900354 and CMMI-1068473. The
authors thank Ohad Perry, the associate editor and two anonymous
referees for their help in the presentation of results and certain
technical issues. 

\bibliographystyle{apacite}
\bibliography{paper2}

\clearpage

\section{Appendix (Intended for an Online Companion)}

In this appendix we prove Theorems 1--3 and provide a brief
description of the experiment design used to build stochastic kriging
metamodels. 

To be self-contained, we first state some definitions, lemmas and
theorems that are used in the proofs. Let $\stackrel{D}\rightarrow$
denote convergence in distribution.
\begin{itemize}

\item \textbf{Borel-Cantelli Lemma} \citep{Billingsley_1995}: For
events $A_1, A_2, \ldots$, if
$\sum_{n=1}^\infty \Pr(A_n)$ converges, then 
\[
\Pr\left(\limsup_n A_n\right)=0 
\]
where 
\[
\limsup_n A_n = \cap_{n=1}^\infty \cup_{k=n}^\infty A_k
\]
is the set of outcomes that occur infinitely many times.

\item \textbf{Lemma 2.11} \citep{Vaart_1998}: Suppose that $\mathbf{X}_n
\stackrel{D}\rightarrow \mathbf{X}$ for a random vector $\textbf{X}$ with a continuous
distribution function. Then the distribution function
of $\textbf{X}_n$ converges uniformly to that of $\textbf{X}$: $\parallel
F_{\textbf{X}_n}-F_\textbf{X}\parallel_{\infty}\rightarrow 0$, where $\parallel
h\parallel_{\infty}$ is the sup-norm of $h$ on $\Re$,
$\parallel h\parallel_{\infty}=\sup_t|h(t)|$.

\item \textbf{Portmanteau Lemma} \citep{Vaart_1998}: For any random
vectors $\mathbf{X}_n$ and $\mathbf{X}$ the following statements are equivalent.
\begin{enumerate}

\item $\mathbf{X}_n\stackrel{D}\rightarrow \mathbf{X}.$

\item $\mbox{E}[f(\mathbf{X}_n)]\rightarrow\mbox{E}[f(\mathbf{X})]$ for all bounded,
continuous functions $f$.

\end{enumerate}

\item \textbf{Theorem 2.3} \citep{Vaart_1998}:  Let $g: \Re^k \rightarrow \Re^m$ be continuous at
every point in a set $\mathcal{C}$ such that $\Pr\{X \in \mathcal{C}\}
= 1$. Then
\begin{enumerate}

\item If $X_n \stackrel{D}{\longrightarrow} X$ then
$g(X_n) \stackrel{D}{\longrightarrow} g(X)$.

\item If $X_n \stackrel{P}{\longrightarrow} X$ then
$g(X_n) \stackrel{P}{\longrightarrow} g(X)$.

\item If $X_n \stackrel{a.s.}{\longrightarrow} X$ then
$g(X_n) \stackrel{a.s.}{\longrightarrow} g(X)$.

\end{enumerate}
In the proofs when we refer to the ``continuous mapping theorem'' we
will mean Theorem 2.3.

\item \textbf{Glivenko-Cantelli Theorem} \citep{Vaart_1998}: If $X_1,X_2,\ldots,X_n$ are i.i.d. random variables with distribution function $F$ and $F_n$ is the empirical cdf of $X_1,X_2,\ldots,X_n$, then $\parallel F_n-F\parallel_{\infty}\stackrel{a.s.}{\longrightarrow}0$ as $n\rightarrow\infty$.

\item \textbf{Lemma 21.2} \citep{Vaart_1998}: For cdf $F$, define the inverse cdf to be
\[
F^{-1}(p) = \inf \{ t{:}\ F(t) \ge p\}.
\]
Then a sequence of cdfs $F_n(t) \rightarrow F(t)$ for every $t$ where
$F$ is continuous if and only if $F_n^{-1}(p) \rightarrow F^{-1}(p)$
for every $p$ where $F^{-1}$ is continuous.

\item \textbf{Theorem~13.1} \citep{Severini_2005}: Let
$\mathbf{X}_1,\mathbf{X}_2,\ldots$ denote a sequence of
$d$-dimensional random vectors such that, for some vector $\bmu$,
\[
\sqrt{n}(\mathbf{X}_n-\bmu)\stackrel{D}\rightarrow \mbox{N}(\mathbf{0}_{d\times
1},\Sigma) \mbox{  as  } n\rightarrow\infty , 
\]
where $\Sigma$ is a $d \times d$ positive definite matrix with $|\Sigma|<\infty$.
Let $g{:}\ \Re^d\rightarrow\Re^k$ denote a continuously differentiable
function and let $\nabla g(\mathbf{x})$ denote the $d\times k$
matrix of partial derivatives of $g$ with respect to $\mathbf{x}$.
Then
\[ 
\sqrt{n}(g(\mathbf{X}_n)-g(\bmu))\stackrel{D}\rightarrow
\mbox{N}(\mathbf{0}_{k\times 1},\nabla g(\bmu)^\top\Sigma
\nabla g(\bmu) )\mbox{  as  } n\rightarrow \infty.  
\] 

\begin{sloppypar}

\item \textbf{Theorem~3.8} \citep{Shao_1995}: Let
$\mathbf{X}_1,\mathbf{X}_2,\ldots,\mathbf{X}_m$ denote $d$-dimensional
i.i.d.\ random vectors and
$\bar{\mathbf{X}}_m=m^{-1}\sum_{i=1}^m\mathbf{X}_i$. Let
$\bar{\mathbf{X}}_m^*=m^{-1}\sum_{i=1}\mathbf{X}^*_i$ where
$\{\mathbf{X}_1^*, \mathbf{X}_2^*, \ldots,\mathbf{X}_m^*\}$ are
randomly and independently drawn with replacement from
$\{\mathbf{X}_1,\mathbf{X}_2, \ldots,\mathbf{X}_m\}$. Let
$g{:}\ \Re^d\rightarrow\Re^k$ denote a continuously differentiable
function and $\nabla g(\bx)$ denote the $d\times k$ matrix of
partial derivatives of $g$ with respect to $\bx$. Let
$T_m=g(\bar{\mathbf{X}}_m)$ and denote the bootstrap variance
estimator for $T_m$ by $v_m^*=\mbox{Var}_*[g(\bar{\mathbf{X}}_m^*)]$.

Suppose that $\mbox{E}[ \mathbf{X}_1^\top \mathbf{X}_1 ]<\infty$ and
$\nabla g(\bmu)\neq \mathbf{0}_{d\times k}$ where
$\bmu=\mbox{E}[\mathbf{X}_1]$.  Suppose further that
\begin{equation}
\label{Condition}
\max_{i_1,\ldots,i_m}|T_m(\mathbf{X}_{i_1},\ldots,
\mathbf{X}_{i_m})-T_m|/\tau_m\stackrel{a.s.}
\rightarrow 0, 
\end{equation}
where the maximum is taken over all integers $i_1,\ldots,i_m$
satisfying $1\leq i_1\leq \cdots\leq i_m\leq m$, and $\{\tau_m\}$ is a
sequence of positive numbers satisfying $\lim\inf_m\tau_m>0$ and
$\tau_m=O(e^{m^q})$ with a $q\in(0,1/2).$ Then $v_m^*$ is strongly
consistent, i.e., $v_m^*/\sigma_m^2\stackrel{a.s.}\rightarrow 1$, where
$\sigma_m^2=m^{-1}\nabla g(\bmu)^\top\Sigma \nabla
g(\bmu)$ and $\Sigma=\mbox{Var}(\mathbf{X}_1).$ 

\end{sloppypar}

\item \textbf{Theorem~1.1} (\cite{Lehmann_Casella_1998},
Chapter~6): Let $X_1,X_2,\ldots,X_m$ be i.i.d. with $\mbox{E}(X_1)=\mu$,
$\mbox{Var}(X_1)=\sigma^2$, and finite fourth moment, and suppose $h$
is a function of a real variable whose first four derivatives
$h^\prime(x),h^{\prime\prime}(x),h^{(3)}(x)$ and $h^{(4)}(x)$ exist for all $x\in I$, where
$I$ is an interval with $\Pr(X_1\in I)=1$. Furthermore, suppose
that $|h^{(4)}(x)|\leq M$ for all $x\in I$, for some $M<\infty$. Then
\[ 
\mbox{E}[h(\bar{X})]=h(\mu)+\frac{\sigma^2}{2m}h^{\prime\prime}(\mu)+\mathcal{R}_m .
\]
If, in addition, the fourth derivative of $h^2$ is also bounded, then
\[ 
\mbox{Var}[h(\bar{X})]=\frac{\sigma^2}{m}[h^\prime(\mu)]^2+\mathcal{R}_m. 
\]
In both cases the remainder $\mathcal{R}_m$ is $O(1/m^2)$.

%, that is, there exist $n_0$ and $A<\infty$ such that $R_n(\mu)<A/n^2$
% for $n>n_0$ and all $\mu$.

\item \textbf{Multivariate Taylor Formula} (\cite{Serfling_2002}, 
page~44): Let the function $g$ defined on $\Re^d$ posses continuous
partial derivatives of order $n$ at each point of an open set
$S\subset \Re^d$. Let $\bx\in S$. For each point $\mathbf{y}$,
$\mathbf{y}\neq \bx$, such that the line segment $L(\bx,\mathbf{y})$
joining $\bx$ and $\mathbf{y}$ lies in $S$, there exists a point
$\mathbf{z}$ in the interior of $L(\mathbf{x},\mathbf{y})$ such that
\begin{eqnarray}
\lefteqn{
g(\mathbf{y})=g(\bx)+ \left. \sum_{k=1}^{n-1}\frac{1}{k!}\sum_{i_1=1}^d\cdots\sum_{i_k=1}^d
\frac{\partial^kg(t_1,\ldots,t_d)}{\partial t_{i_1}\cdots \partial
t_{i_k}}
\right|_{\mathbf{t}=\bx}\cdot \prod_{j=1}^k(y_{i_j}-x_{i_j}) }
\nonumber \\
&& + \left.\frac{1}{n!}\sum_{i_1=1}^d\cdots\sum_{i_n=1}^d\frac{\partial^n
g(t_1,\ldots,t_d)}{\partial t_{i_1}\cdots \partial t_{i_n}}
\right|_{\mathbf{t}=\mathbf{z}}\cdot\prod_{j=1}^n(y_{i_j}-x_{i_j}). \nonumber
\end{eqnarray}

\begin{comment}
\textcolor{black}{
\item \textbf{Theorem~2.1} in Chapter~3 \citep{Stein_Shakarchi_2005}: Let $K_{\delta}$ denote a Gaussian kernel with variance equal $\delta^2>0$: $K_{\delta}(x)=\frac{1}{\sqrt{2\pi}\delta}\exp^{-x^2/2\delta^2}$. A function $f$ is integrable on $\Re^d$ and let 
\[ (f\ast K_{\delta})(x)=\int_{\Re^d}f(x-y)K_{\delta}(y)dy.\] Then
\[(f\ast K_{\delta})(x)\rightarrow f(x) \mbox{ as } \delta\rightarrow 0 \]
for every $x$ in the Lebesgue set of $f$. In particular, the limit holds for a.e. $x$. 
}
\end{comment}

\end{itemize}

\subsection{Asymptotic Consistency of $\mbox{CI}_+$}

To prove Theorem~1, we first establish three supporting lemmas.

\begin{lemma}
\label{lemma1}
Suppose that Assumptions~1--2 hold. Then
the bootstrap resampled moments converge almost surely to the true
moments $\widehat{\mathbf{X}}_m\stackrel{a.s.}\rightarrow
\mathbf{x}_c$ as $m\rightarrow\infty$.
\end{lemma}

%\vspace{12pt}

\noindent \textbf{Proof:} Since all of the input processes are
independent, we establish the result for one input distribution $F^c$
without loss of generality. We prove the result for $\mathbf{x}_c$
being the generic $h$th-order moment, $\alpha_h \equiv \mbox{E}(Z^h) <
\infty$, for $Z\sim F^c$. 

The $h$th-order bootstrap resampled moment is
\begin{equation}
\label{eq:boot.moment}
\widehat{X}_{m}=\frac{1}{m}\sum_{j=1}^m (\textcolor{black}{Z^{(j;m)}})^h \mbox{ with }
\textcolor{black}{Z^{(j;m)}}\stackrel{i.i.d}\sim  \mathbf{Z}_{m}^{(0)}  
\end{equation}
where ``$\textcolor{black}{Z^{(j;m)}}\sim \mathbf{Z}_{m}^{(0)}$"
denotes the $j$th
independent sample with replacement from $\mathbf{Z}_{m}^{(0)}$. We
use the Chebychev Inequality and the Borel-Cantelli Lemma to prove the
result.

By the Chebychev Inequality, for every $\epsilon>0$, we have
\begin{equation}
\label{eq.ChebIneq}
 \mbox{Pr}\left\{|\widehat{X}_m-\alpha_h|>\epsilon \right\}
\leq
\frac{\mbox{E}\left[ (\widehat{X}_m-\alpha_h)^4 \right]}{\epsilon^4}. 
\end{equation}
Notice that
\begin{equation}
\mbox{E}\left[(\widehat{X}_m-\alpha_h)^4\right]
=\mbox{E}\left[\widehat{X}_m^4\right]-4\alpha_h\mbox{E}\left[\widehat{X}_m^3\right]
+6\alpha_h^2\mbox{E}\left[\widehat{X}_m^2\right]-4\alpha_h^3\mbox{E}\left[\widehat{X}_m\right]+\alpha_h^4.
\label{eq.fourM}
\end{equation}
We will analyze each term in Equation~(\ref{eq.fourM}). First, we show
that any $i$th bootstrap resampled moment, denoted as
$\widehat{\alpha}_i$, is unbiased,
\begin{eqnarray}
\label{eq.unbiasedM}
\lefteqn{ \mbox{E}\left[ \widehat{\alpha}_i \right]\equiv\mbox{E}\left[\frac{1}{m}\sum_{j=1}^m(\textcolor{black}{Z^{(j;m)}})^i
\right]} \\ 
&=& \mbox{E}\left[\mbox{E}\left[(\textcolor{black}{Z^{(j;m)}})^i|Z_1^{(0)},\ldots,
Z_m^{(0)}\right]\right] \nonumber \\ 
&=& \mbox{E}\bigg[\frac{1}{m}\sum_{j=1}^m (Z_j^{(0)})^i\bigg]   \nonumber \\ \nonumber
&=& \alpha_i.  \nonumber
\end{eqnarray}
Thus, $\mbox{E}[\widehat{X}_m]=\alpha_h$. Notice that
\begin{eqnarray}
\lefteqn{\mbox{E}\left[ \widehat{X}_m^2 \right]= \mbox{E}\left[\left(\frac{1}{m}\sum_{j=1}^m(\textcolor{black}{Z^{(j;m)}})^h\right)^2\right] } \nonumber \\ \nonumber
&=& \frac{1}{m^2}\mbox{E}\left[\left(\sum_{j=1}^m(\textcolor{black}{Z^{(j;m)}})^h\right)^2\right] \nonumber \\ 
&=& \frac{1}{m^2}\mbox{E}\left[\sum_{j=1}^m(\textcolor{black}{Z^{(j;m)}})^{2h}
+\sum_{i\neq j}(\textcolor{black}{Z^{(i;m)}})^h(\textcolor{black}{Z^{(j;m)}})^h\right]  \nonumber \\
&=&  \frac{1}{m^2}\left(m\alpha_{2h}+m(m-1) \mbox{E}\left[\mbox{E}[(\textcolor{black}{Z^{(i;m)}})^h | Z_1^{(0)},\ldots,Z_m^{(0)}]\cdot \mbox{E}[(\textcolor{black}{Z^{(j;m)}})^h |Z_1^{(0)},\ldots,Z_m^{(0)}]\right] \right) \nonumber \\ 
&=&  \frac{1}{m^2}\left(m\alpha_{2h}+
m(m-1)
\mbox{E}\bigg[\bigg(\frac{1}{m}\sum_{i=1}^m (Z_i^{(0)})^h\bigg)^2\bigg]\right)  \nonumber\\ 
&=&   \frac{1}{m^2}\left(m\alpha_{2h}+
\frac{m(m-1)}{m^2} \mbox{E}\bigg[\sum_{i=1}^m(Z_i^{(0)})^{2h} + \sum_{i\neq j}(Z_i^{(0)})^h(Z_j^{(0)})^h\bigg]\right)  \nonumber\\ 
&=&  \frac{1}{m^2}\left(m\alpha_{2h}+
\frac{m(m-1)}{m^2}(m\alpha_{2h}+m(m-1)\alpha_h^2) \right)  \nonumber\\
&=& \frac{1}{m^2}[(2m-1)\alpha_{2h}+(m-1)^2\alpha_h^2] \nonumber\\
&=&\frac{2}{m}\alpha_{2h}+\left(1-\frac{2}{m}\right)\alpha_h^2+O(m^{-2}) \nonumber
\end{eqnarray}
where $O(m^{-2})$ means terms at most order $1/m^2$. 
Similar derivations show that 
\begin{eqnarray}
\lefteqn{\mbox{E}\left[\widehat{X}_m^3\right]
=\frac{1}{m^4}\Big([m(4m-3)+(m-1)(m-2)]\alpha_{3h}} \nonumber\\ 
&& +\ [3m(m-1)^2+3(m-1)^2(m-2)]\alpha_h\alpha_{2h}
+(m-1)^2(m-2)^2\alpha_h^3\Big)  \nonumber\\ 
&=& \frac{6}{m}\alpha_h\alpha_{2h}+\left(1-\frac{6}{m}\right)\alpha_h^3+O(m^{-2}) \nonumber 
\end{eqnarray}
and 
\[
\mbox{E}\left[\widehat{X}_m^4\right]=\frac{12}{m}\alpha_h^2\alpha_{2h}
+\left(1-\frac{12}{m}\right)\alpha_h^4+ O(m^{-2}).
\]
Thus,
\begin{eqnarray}
\label{eq.FourCentM}
\lefteqn{
\mbox{E}\left[(\widehat{X}_m-\alpha_h)^4\right] 
= \mbox{E}[\widehat{X}_m^4]-4\alpha_h\mbox{E}[\widehat{X}_m^3]
+6\alpha_h^2\mbox{E}[\widehat{X}_m^2]-4\alpha_h^3\mbox{E}[\widehat{X}_m]
+\alpha_h^4} \\
&=& 
\frac{12}{m}\alpha_h^2\alpha_{2h}
+\left(1-\frac{12}{m}\right)\alpha_h^4
-4\alpha_h\left[\frac{6}{m}\alpha_h\alpha_{2h}
+\left(1-\frac{6}{m}\right)\alpha_h^3\right]
\nonumber\\ 
&& +\ 6\alpha_h^2\left[\frac{2}{m}\alpha_{2h}+\left(1-\frac{2}{m}\right)\alpha_h^2\right]
-3\alpha_h^4+O(m^{-2}) \nonumber\\ 
&=& 0+ O(m^{-2}) \nonumber
\end{eqnarray}
because all of the $O(m^{-1})$ terms cancel.
Therefore, combining Equations~(\ref{eq.ChebIneq}), (\ref{eq.fourM})
and (\ref{eq.FourCentM}), we have
\[ \sum_{m=1}^{\infty}\mbox{Pr}\{|\widehat{X}_m-\alpha_h|>\epsilon\}\leq 
\sum_{m=1}^{\infty}\frac{c}{m^2\epsilon^4} < \infty\] where $c$ is
some finite constant.  Thus, if $\alpha_{4h} <\infty$, then
$\widehat{X}_m\stackrel{a.s.}\rightarrow \alpha_h$ by the
first Borel-Cantelli Lemma in Section~4 of \cite{Billingsley_1995}. 

Since Assumption~2 guarantees $m_\ell\rightarrow\infty$ for each
moment associated with the $\ell$th input distribution, we can generalize
the almost sure convergence to a vector of moments by applying the
converging together lemma. Therefore, we have
$\widehat{\mathbf{X}}_{\mathbf{m}}\stackrel{a.s.}\rightarrow
\mathbf{x}_c$. \blot

\vspace{12pt}

\noindent \textbf{Remark:} The independent variables in our stochastic
kriging metamodel consist of central moments and standardized central
moments.  Since standardized moments are continuous functions of raw
moments, we can use the continuous mapping theorem to obtain
corresponding almost sure convergence of the standardized moments.

\vspace{12pt}

% ========================================================
\begin{sloppypar} 
Given a fixed and finite number of design points
$\bx_1,\bx_2,\ldots,\bx_k$, let
$\mathbf{M}=(M(\bx_1),M(\bx_2),\ldots,M(\bx_k))^\top$.  The simulation
error at design point $\bx_i$ is $\epsilon(\bx_i)$, so let
$\bar{\epsilon}(\bx_i)=\sum_{j=1}^{n_i}\epsilon(\bx_i)/n_i$ for
$i=1,2,\ldots,k$ denote the average. Therefore, the sample means of
simulation outputs at all design points can be represented as
$\bar{\mathbf{Y}}_\mathcal{D}=\mathbf{M}+\bar{\boldmath{\epsilon}}$,
where $\bar{\boldmath{\epsilon}}=(\bar{\epsilon}(\bx_1),
\bar{\epsilon}(\bx_2),\ldots,\bar{\epsilon}(\bx_k))^\top$. Finally,
let $M_p(\cdot)$ be a GP having the conditional distribution of
$M(\cdot)$ given $\bar{\mathbf{Y}}_\mathcal{D}$. 
\end{sloppypar}

\vspace{12pt}

\begin{lemma}
\label{lemma2}
Suppose Assumptions~3--4 hold. Then
$M_p(\cdot)$ has continuous sample paths almost surely.
\end{lemma}

\noindent \textbf{Proof:} Let $(\Omega_M, P_M)$ be the underlying
probability space for the GP $M(\cdot)$, and $(\Omega_\epsilon, P_\epsilon)$
be the underlying probability space for $\bar{\boldmath{\epsilon}}$. Notice
that $(\Omega_\epsilon, P_\epsilon)$ depends on the particular design
points $\bx_1, \bx_k, \ldots, \bx_k$ and corresponding numbers of
replications $n_1, n_2, \ldots, n_k$ which we consider fixed and
given, while $(\Omega_M, P_M)$ does not. 

Let $\omega_M \in \Omega_M$ be an elementary outcome and $M(\cdot,
\omega_M)$ the resulting random function. For notational convenience,
let $\mathbf{M}(\omega_M) = \left( M(\bx_1, \omega_M), M(\bx_2,
\omega_M), \ldots, M(\bx_k, \omega_M) \right)^\top$ the random
function evaluated at $\bx_1, \bx_2, \ldots, \bx_k$. Similarly,
$\bar{\boldmath{\epsilon}} =
\bar{\boldmath{\epsilon}}(\omega_\epsilon)$ for elementary outcome
$\omega_\epsilon \in \Omega_\epsilon$. Notice that under Assumption~3,
$\bar{\boldmath{\epsilon}}(\omega_\epsilon)$ has a multivariate normal
distribution.

Theorem 3.4.1 of \cite{Adler_2010} asserts that there is a
$P_M$-measurable set $\Omega_M^c \subset \Omega_M$ such that $\Pr\{
\omega_M \in \Omega_M^c\} = P_M(\Omega_M^c) = 1$, and for every
$\omega_M \in \Omega_M^c$ the function $M(\cdot, \omega_M)$ is
continuous.

%We can partition the sample space of the Gaussian process as $\Omega_M
%= \Omega_M^c \cup \bar{\Omega}_M^c$ where $\Omega_M^c$ are all
%outcomes for which $M(\cdot, \omega_M)$ is continuous; under our
%assumptions $P_M(\Omega_M^c) = 1$ and thus $P_M(\bar{\Omega}_M^c) = 0$.

The random variable $\bar{\bY}_\mathcal{D}$ maps $\Omega = \Omega_M
\times \Omega_\epsilon \rightarrow \Re^k$ as
$\bar{\bY}_\mathcal{D}(\omega) = \mathbf{M}(\omega_M) +
\bar{\mathbf{\epsilon}}(\omega_\epsilon)$ for $\omega = (\omega_M,
\omega_\epsilon) \in \Omega$ with probability measure $P = P_M \cdot
P_\epsilon$ since they are independent.  Our goal is to prove that
\begin{equation}
\label{eq:Prob0}
\Pr\{\omega_M \in \bar{\Omega}_M^c | \bar{\bY}_{D}\} = 0
\end{equation}
almost surely.

We know that $0 \le \Pr\{\omega_M \in \bar{\Omega}_M^c | \bar{\bY}_{D}\} \le 1$
with probability $1$. But also
\[
0 = \Pr\{\omega_M \in \bar{\Omega}_M^c \} = 
\E\left[\Pr\{\omega_M \in \bar{\Omega}_M^c | \bar{\bY}_{D}\} \right] .
\]
Therefore, (\ref{eq:Prob0}) must hold. \blot

\vspace{12pt}

% ========================================================

\begin{lemma}
\label{lemma3}
Suppose that Assumptions~1--4 hold.  Then
$M_p(\widehat{\mathbf{X}}_{\mathbf{m}})\stackrel{a.s.} \rightarrow
M_p(\mathbf{x}_c)$ as $m\rightarrow\infty$.
\end{lemma} 

\noindent \textbf{Proof:} Under Assumption~3, the GP $M(\cdot)$ has
continuous sample paths almost surely; applying Lemma~2, $M_p(\cdot)$
also has continuous sample paths almost surely.  Under
Assumptions~1--2,
$\widehat{\mathbf{X}}_{\mathbf{m}}\stackrel{a.s.}\rightarrow\mathbf{x}_c$
as $m\rightarrow \infty$ by Lemma~1. And $M_p(\cdot)$
and $\widehat{\mathbf{X}}_{\mathbf{m}}$ are independent. The result
follows by applying the continuous mapping theorem. \blot

\vspace{12pt}

% ================================================================

\setcounter{theorem}{0}
\begin{theorem}
Suppose that Assumptions 1--4 hold.  Then the interval $[M_{(\lceil
B\frac{\alpha}{2} \rceil)},M_{(\lceil B(1-\frac{\alpha}{2})\rceil)}]$
is asymptotically consistent, meaning
\begin{equation}
\lim_{m\rightarrow\infty} \lim_{B\rightarrow \infty} 
\Pr\{M_{(\lceil B\alpha/2 \rceil)} \leq M_p(\mathbf{x}_c) \leq M_{(\lceil B(1-\alpha/2)\rceil)} \} = 1-\alpha.
\label{eq.11}
\end{equation} 
\end{theorem}

\noindent \textbf{Proof:} Define $K_{\mathbf{m}}(t)\equiv
\Pr\left\{ M_p(\widehat{\mathbf{X}}_{\mathbf{m}})\leq t \right\}. $ Notice
that the distribution $K_{\mathbf{m}}(t)$ depends on both the
distributions of $ M_p(\cdot)$ and
$\widehat{\mathbf{X}}_{\mathbf{m}}$. Specifically,
\begin{eqnarray*}
K_{\mathbf{m}}(t) &=& 
	\int \Pr\left\{M_p(\bx) \le t | \widehat{\bX}_\bm = \bx\right\}
	d\widehat{F}_{\bX_\bm}(\bx |\bz_\bm^{(0)}) \\
&=&
\int \Phi\left(
\frac{t - m_p(\bx)}{\sigma_p(\bx)}
\right)
	d\widehat{F}_{\bX_\bm}(\bx |\bz_\bm^{(0)}).
\end{eqnarray*}
Thus, $K_{\mathbf{m}}(t)$ is a continuous distribution almost surely.
Let $\widehat{K}_\bm$ be the empirical cdf of $M_1, M_2, \ldots, M_B$,
which are i.i.d.\ from $K_{\mathbf{m}}(t)$.  Notice that
$\widehat{K}_\bm^{-1}(\gamma) = M_{(\lceil B\gamma\rceil)}$ for
$\gamma = \alpha/2$ and $1 - \alpha/2$.

By the Glivenko-Cantelli Theorem \citep{Vaart_1998}, $||K_\bm -
\widehat{K}_\bm||_\infty \stackrel{a.s.}{\longrightarrow} 0$ as $B
\rightarrow \infty$. 
Therefore, by Lemma~21.2 of \cite{Vaart_1998}, 
\[
| M_{(\lceil B\gamma\rceil)} - K_\bm^{-1}(\gamma) |
\stackrel{a.s.}{\longrightarrow} 0
\]
as $B \rightarrow \infty$ for $\gamma = \alpha/2, 1 - \alpha/2$. As a result,
\[
\lim_{B\rightarrow \infty} 
\Pr\{M_{(\lceil B\alpha/2 \rceil)} \leq M_p(\mathbf{x}_c) \leq
M_{(\lceil B(1-\alpha/2)\rceil)} \} = 
\Pr\{K_\bm^{-1}(\alpha/2) \leq M_p(\mathbf{x}_c) \leq 
K_\bm^{-1}(1-\alpha/2) \}.
\]
Therefore, Equation~(\ref{eq.11}) becomes
\begin{equation}
\lim_{m\rightarrow\infty}\Pr\{K_{\mathbf{m}}^{-1}(\alpha/2)\leq
M_p(\mathbf{x}_c)  \leq K_{\mathbf{m}}^{-1}(1-\alpha/2)\}=1-\alpha.
\label{eq.2}
\end{equation}

To show Equation~(\ref{eq.2}), we only need to show that
\begin{equation}
\lim_{m\rightarrow \infty} \Pr\{K_{\mathbf{m}}^{-1}(\alpha/2) > M_p(\mathbf{x}_c)\}=\alpha/2 \nonumber
\label{eq.3}
\end{equation}
because the proof of the upper bound is similar.

%As $B\rightarrow\infty$, $M_{(\lceil B\alpha/2 \rceil)}$ and
%$M_{(\lceil B(1-\alpha/2)\rceil)}$ converge to the true bootstrap
%quantiles of $M_p(\widehat{\mathbf{X}}_{\mathbf{m}})$. Thus,
%Equation~(\ref{eq.11}) becomes

Since, conditional on $\bar{\bY}_\mathcal{D}$, $M_p(\mathbf{x}_c)\sim
\mbox{N}(m_p(\mathbf{x}_c),\sigma_p^2(\mathbf{x}_c))$, the cdf $H(t)\equiv
\Pr\{M_p(\mathbf{x}_c)\leq t\}$ is continuous. By Lemma~3 and
Lemma~2.11 in \cite{Vaart_1998},
\begin{eqnarray}
\lefteqn{\sup_t |\Pr\{M_p(\widehat{\mathbf{X}}_{\mathbf{m}})\leq t \}-\Pr\{M_p(\mathbf{x}_c)\leq t\} |} \nonumber\\ 
&=&\parallel K_{\mathbf{m}}-H \parallel_{\infty}  \rightarrow 0 \mbox{ as } m\rightarrow \infty. \nonumber
\end{eqnarray}
 
Therefore, 
\begin{eqnarray}
\lefteqn{ \Pr\{K_{\mathbf{m}}^{-1}(\alpha/2)>M_p(\mathbf{x}_c)  \}}  \nonumber\\ 
&=& \Pr\{\alpha/2 \geq K_{\mathbf{m}}(M_p(\mathbf{x}_c))  \}  \nonumber\\ 
\label{eq.temp1}
&=& \Pr\{\alpha/2 \geq H(M_p(\mathbf{x}_c))\} + \textcolor{black}{o(1)}  \\ 
&=& \Pr\{M_p(\mathbf{x}_c)\leq H^{-1}(\alpha/2) \} +\textcolor{black}{o(1)} \nonumber\\ 
&=& \alpha/2 + \textcolor{black}{o(1)}. \nonumber
\end{eqnarray}
Equation~(\ref{eq.temp1}) is obtained because 
\[ |K_{\mathbf{m}}(M_p(\mathbf{x}_c))-H(M_p(\mathbf{x}_c))|\leq \parallel K_{\mathbf{m}}-H\parallel_{\infty}\rightarrow 0 \mbox{  as $m\rightarrow \infty$}. \]
Thus, we have
\begin{eqnarray}
\lefteqn{ \lim_{m\rightarrow\infty} \Pr\{ K_{\mathbf{m}}^{-1}(\alpha/2) \leq M_p(\mathbf{x}_c) \leq K_{\mathbf{m}}^{-1}(1-\alpha/2)\} }   \nonumber\\ 
&=& \lim_{m\rightarrow\infty} \Pr \{M_p(\mathbf{x}_c)\leq K_{\mathbf{m}}^{-1}(1-\alpha/2)\}-\lim_{m\rightarrow\infty} \Pr\{M_p(\mathbf{x}_c)<K_{\mathbf{m}}^{-1}(\alpha/2)\}  \nonumber\\ 
&=& (1-\alpha/2)-\alpha/2  \nonumber\\ 
&=& 1-\alpha. \nonumber
\end{eqnarray} \blot

\vspace{12pt}

% =================================================================

\subsection{Asymptotic Analysis of Variance Component Estimators}

\begin{theorem}
Suppose that Assumptions~1--4 hold. Then
the variance component estimators $\widehat{\sigma}^2_M,
\widehat{\sigma}^2_I, \widehat{\sigma}^2_T$ are consistent as 
$m, B\rightarrow\infty$. 
\end{theorem}

%Specifically, 
%\begin{itemize} 
%\item As
%$m\rightarrow\infty$, the input uncertainty disappears:
%\[
%\lim_{m\rightarrow\infty}\sigma^2_M=\sigma^2_{p}(\bx_c),
%\lim_{m\rightarrow\infty}\sigma^2_I=0 \mbox{ and }
%\lim_{m\rightarrow\infty}\sigma^2_T=\sigma^2_{p}(\bx_c).
%\] 

%\item As
%$m\rightarrow\infty$ and $B\rightarrow\infty$, the variance component
%estimators are consistent: 
%\[
%\lim_{m\rightarrow\infty}\lim_{B\rightarrow\infty}\widehat{\sigma}_M^2
%=\lim_{m\rightarrow\infty}\sigma_M^2=\sigma_p^2(\bx_c), 
%\]
%\[
%\lim_{m\rightarrow\infty}\lim_{B\rightarrow\infty}\widehat{\sigma}_I^2
%=\lim_{m\rightarrow\infty}\sigma_I^2=0, 
%\] 
%\[
%\lim_{m\rightarrow\infty}\lim_{B\rightarrow\infty}\widehat{\sigma}_T^2
%=\lim_{m\rightarrow\infty}\sigma_T^2=\sigma^2_p(\bx_c). 
%\]
%\end{itemize}

\noindent \textbf{Proof:} When a GP $M(\cdot)$ has a continuous
correlation function with all parameters finite, the SK predictor
\begin{equation}
m_{p}(\bx)=\widehat{\beta}_0
+\tau^2R(\bx)^\top[\Sigma+C]^{-1}(\bar{\bY}_\mathcal{D}-\widehat{\beta}_0\cdot 1_{k\times 1}),
\end{equation}
and corresponding variance
\begin{equation}
\sigma_{p}^2(\bx) = \tau^2-\tau^4R(\bx)^\top[\Sigma+C]^{-1}R(\bx) 
 +\mathbf{\eta}^\top[1_{k\times 1}^\top(\Sigma+C)^{-1}1_{k\times 1}]^{-1}\mathbf{\eta}  \nonumber
\end{equation}
where $R(\bx)^\top=(r(\bx-\bx_1),r(\bx-\bx_2),\ldots,r(\bx-\bx_k))$ and $\eta=1-1_{k\times 1}^\top(\Sigma+C)^{-1}\tau^2R(\bx)$, are continuous and bounded functions of $\bx$. 

By the Strong Law of Large Numbers, the raw moment estimator
$\mathbf{X}_{\mathbf{m}}\stackrel{a.s.}\rightarrow\bx_c$ as
$m\rightarrow \infty$ under Assumptions~1--2. This almost sure
convergence can be extended to central moments and standardized
central moments by the continuous mapping theorem. By applying
the Portmanteau Lemma in \cite{Vaart_1998}, we have 
\[\lim_{m\rightarrow\infty}\sigma_M^2
=\lim_{m\rightarrow\infty} \int \sigma^2_p(\mathbf{x})\, dF^c_{\mathbf{X}_{\mathbf{m}}}(\mathbf{x})
=\lim_{m\rightarrow\infty}\mbox{E}[\sigma^2_p(\mathbf{X}_{\mathbf{m}})]
=\sigma_p^2(\bx_c),  \]
and
\begin{eqnarray}
\lefteqn{\lim_{m\rightarrow\infty}\sigma_I^2
=\lim_{m\rightarrow\infty}\int(m_p(\mathbf{x})-\mu_0)^2\, dF^c_{\mathbf{X}_{\mathbf{m}}}(\mathbf{x})} \nonumber\\ 
&=& \lim_{m\rightarrow\infty}\mbox{E}\left[\big(m_p(\mathbf{X}_{\mathbf{m}})
-\mbox{E}[m_p(\mathbf{X}_{\mathbf{m}})]\big)^2\right] \nonumber\\ 
&=& \big(m_p(\bx_c)-m_p(\bx_c)\big)^2=0  \nonumber
\end{eqnarray} 
where $\mu_0=\int\int \nu \, dF(\nu|\mathbf{x})\,
dF^c_{\mathbf{X}_{\mathbf{m}}}(\mathbf{x})=\int m_p(\mathbf{x})\,
dF^c_{\mathbf{X}_{\mathbf{m}}}(\mathbf{x})$.

Recall that $F(\nu|\bx)$ is a normal distribution
$\mbox{N}(m_p(\bx),\sigma_p^2(\bx))$. Let $g(\bx)\equiv
\int(\nu-\mu_0)^2\, dF(\nu|\bx)$. 
Then 
\begin{eqnarray}
\lefteqn{\lim_{m\rightarrow\infty}\sigma_T^2 = 
\lim_{m\rightarrow\infty}
 \int\int (\nu-\mu_0)^2\, dF(\nu|\mathbf{x})\, dF^c_{\mathbf{X}_{\mathbf{m}}}(\mathbf{x}) } 
 \nonumber\\[6pt]
&=&
\lim_{m\rightarrow\infty}
 \int g(\bx)\, dF^c_{\mathbf{X}_{\mathbf{m}}}(\mathbf{x}).  \nonumber
\end{eqnarray}
However,
\begin{eqnarray*}
g(\bx) &=& 
\int (\nu - m_p(\bx) + m_p(\bx) - \mu_0 )^2\, dF(\nu | \mathbf{x}) \\
&=&
\int (\nu - m_p(\bx))^2\, dF(\nu |\mathbf{x})
 + (m_p(\bx) - \mu_0 )^2  
+ (m_p(\bx) - \mu_0 ) \int (\nu - m_p(\bx))\, dF(\nu | \mathbf{x}) \\
&=&
\sigma_p^2(\bx) + (m_p(\bx) - \mu_0 )^2   + 0.
\end{eqnarray*}
Since $m_p(\bx)$ and $\sigma_p^2(\bx)$ are continuous and bounded
functions, so is $g(\bx)$. Therefore,  
\[
\lim_{m\rightarrow\infty}\sigma_T^2 = 
 \lim_{m\rightarrow\infty}\mbox{E}[g(\mathbf{X}_{\mathbf{m}})]
= g(\bx_c) =\sigma^2_p(\bx_c)
\]
by applying the Portmanteau Lemma.

\begin{comment} % original proof
The function $g(\bx)$ is continuous
and bounded in a finite neighborhood of $\bx_c$. Since
$\mathbf{X}_{\mathbf{m}}\stackrel{a.s.}\rightarrow\bx_c$, there exists
$\tilde{\mathbf{m}}$ such that $\mathbf{X}_{\mathbf{m}}$ is contained
in this neighborhood for all $\mathbf{m}>\tilde{\mathbf{m}}$ (term by
term) almost
surely. By applying the Portmanteau Lemma, we have 
\begin{eqnarray}
\lefteqn{\lim_{m\rightarrow\infty}\sigma_T^2 = 
\lim_{m\rightarrow\infty}
 \int\int (\nu-\mu_0)^2\, dF(\nu|\mathbf{x})\, dF^c_{\mathbf{X}_{\mathbf{m}}}(\mathbf{x}) } 
 \nonumber\\[6pt]
&=& \lim_{m\rightarrow\infty}\mbox{E}[g(\mathbf{X}_{\mathbf{m}})] \nonumber\\ 
&=& g(\bx_c) =\sigma^2_p(\bx_c).  \nonumber
\end{eqnarray}
\end{comment}

Next, we will show consistency of the variance estimators. By Lemma~1, $\widehat{\mathbf{X}}_{\mathbf{m}}\stackrel{a.s.}\rightarrow \bx_c.$
For the metamodel uncertainty estimator, we have
\begin{eqnarray}
\lefteqn{\lim_{m\rightarrow\infty}\lim_{B\rightarrow\infty}\widehat{\sigma}^2_M
=\lim_{m\rightarrow\infty}\lim_{B\rightarrow\infty}\frac{1}{B}
\sum_{b=1}^B\sigma^2_p
(\widehat{\mathbf{X}}_{\mathbf{m}}^{(b)}) }
\nonumber\\ 
&=& \lim_{m\rightarrow\infty}\mbox{E}\Big[\sigma_p^2
(\widehat{\mathbf{X}}_{\mathbf{m}})|\mathbf{Z}_{\mathbf{m}}^{(0)}\Big] 
\nonumber\\ 
&=&\sigma_p^2(\bx_c). \nonumber
\end{eqnarray}
The last step follows by applying the Portmanteau Lemma.

For the input uncertainty estimator, we have
\begin{eqnarray}
\lefteqn{
\lim_{m\rightarrow\infty}\lim_{B\rightarrow\infty}\widehat{\sigma}^2_I
=\lim_{m\rightarrow\infty}\lim_{B\rightarrow\infty}
\frac{B}{B-1}\left[\frac{1}{B}\sum_{b=1}^B m_p^2(
\widehat{\mathbf{X}}_{\mathbf{m}}^{(b)})-\bar{{\mu}}^2\right] } 
\nonumber\\ 
&=& \lim_{m\rightarrow\infty}\bigg(
\mbox{E}\Big[m_p^2(\widehat{\mathbf{X}}_{\mathbf{m}})|\mathbf{Z}_{\mathbf{m}}^{(0)}\Big]
-\mbox{E}^2\Big[m_p(\widehat{\mathbf{X}}_{\mathbf{m}})
|\mathbf{Z}_{\mathbf{m}}^{(0)}\Big] \bigg) \nonumber\\ 
&=& m_p^2(\bx_c)-m_p^2(\bx_c)=0. \nonumber
\end{eqnarray}
The last step follows by applying Lemma~1 and the Portmanteau Lemma.

For the total variance estimator, we have
\begin{eqnarray}
\lefteqn{\lim_{m\rightarrow\infty}\lim_{B\rightarrow\infty}
\widehat{\sigma}^2_T = \lim_{m\rightarrow\infty}\lim_{B\rightarrow\infty} \frac{B}{B-1}\left(\frac{1}{B}\sum_{b=1}^B M_b^2-\bar{M}^2\right)} \nonumber\\ 
&=&\lim_{m\rightarrow\infty}\mbox{E}\big[M_p^2(\widehat{\mathbf{X}}_{\mathbf{m}})
|\mathbf{Z}_{\mathbf{m}}^{(0)}\big] - \lim_{m\rightarrow\infty}\mbox{E}^2\big[M_p
(\widehat{\mathbf{X}}_{\mathbf{m}})|\mathbf{Z}_{\mathbf{m}}^{(0)}\big] 
\nonumber \\ 
\label{eq.Th2Mid1}
&=& \mbox{E}[M_p^2(\bx_c)]-\mbox{E}^2[M_p(\bx_c)]  \\
&=& m_p^2(\bx_c)+\sigma_p^2(\bx_c)-m_p^2(\bx_c) \nonumber\\ 
&=& \sigma^2_p(\bx_c). \nonumber
\end{eqnarray}
By Lemma~3,
$M_p(\widehat{\mathbf{X}}_{\mathbf{m}})\stackrel{a.s.}\rightarrow
M_p(\bx_c) \mbox{ as } m\rightarrow\infty$. Then
Step~(\ref{eq.Th2Mid1}) follows by applying Portmanteau Lemma. \blot

\vspace{12pt}

%===============================================================

\begin{theorem}
Suppose that Assumptions~1--4 and the
following additional assumptions hold:
\setcounter{enumi}{4}
\begin{enumerate}

\item [5.]The first three derivatives of the correlation function of
the GP $M(\bx)$ exist and the third derivative is bounded; and

\item [6.] $m_\ell/m \rightarrow 1$ for $\ell=1,2,\ldots,L$.

\end{enumerate}
Then $\lim_{m\rightarrow\infty}m\sigma_I^2=\lim_{m\rightarrow\infty}
\lim_{B\rightarrow\infty} m\widehat{\sigma}^2_I=\sigma^2_\mu$ almost
surely, where
$\sigma^2_\mu$ is a positive constant.
\end{theorem}

\noindent \textbf{Proof:} 
Under Assumptions~1--2, and applying the multivariate central limit
theorem, we have as $m\rightarrow\infty$,
\[
\sqrt{m}(\mathbf{X}_m-\mathbf{x}_c)\stackrel{D}\rightarrow\mbox{N}(\mathbf{0}_{d\times 1},\Lambda)
\]
where $\Lambda$ denotes the $d\times d$ positive definite  asymptotic
variance-covariance matrix of $\bX_{\mathbf{m}}$.

When a GP
$M(\bx)$ has a continuous correlation function with all parameters
finite, the SK predictor 
\begin{equation}
m_{p}(\bx)=\widehat{\beta}_0
+\tau^2R(\bx)^\top[\Sigma+C]^{-1}
(\bar{\bY}_\mathcal{D}-\widehat{\beta}_0\cdot 1_{k\times 1}),
\label{eq.predictor}
\end{equation} 
given the simulation sample mean $\bar{\bY}_\mathcal{D}$, is continuous and bounded. 
Under Assumption~5, the gradient $\nabla m_p(\bx)$ exists and
is
continuous. We will show that $\nabla m_p(\bx)\neq
\mathbf{0}_{d\times 1}$ almost surely. By taking the derivative of
$m_p(\bx)$ in Equation~(\ref{eq.predictor}), we have
\begin{equation}
 \frac{\partial m_{p}(\bx)}{\partial x_j}=\underbrace{
\frac{\partial{R}(\bx)^\top}{\partial
x_j}\tau^2[\Sigma+C]^{-1}}_\mathbf{A}(\bar{\bY}_\mathcal{D}-\widehat{\beta}_0\cdot
1_{k\times 1}).
\end{equation}
Since $\partial{R}(\bx)^\top/\partial x_j=
(\partial{r}(\bx-\bx_1)/\partial x_j,\partial{r}(\bx-\bx_2)/\partial
x_j,\ldots, \partial{r}(\bx-\bx_k)/\partial x_j)\neq
\mathbf{0}_{1\times k}$ and $\tau^2[\Sigma+C]^{-1}$ is positive
definite, $\mathbf{A}$ is a non-zero constant vector. Under
Assumption~3, $\mathbf{A}(\bar{\bY}_\mathcal{D}-\widehat{\beta}_0\cdot
1_{k\times 1})$ is a normal random variable that is equal to 0 with
probability 0. Thus, $\nabla m_p(\bx)\neq \mathbf{0}_{d\times
1}$ almost surely.  Applying Theorem~13.1 in \cite{Severini_2005}, we have
\[\sqrt{m}(m_p(\mathbf{X}_{{m}})-m_p(\mathbf{x}_c))\stackrel{D}
\rightarrow\mbox{N}(0,\sigma^2_{\mu}) \]
where $\sigma_{\mu}^2=\nabla
m_p(\bx_c)^\top\Lambda\nabla m_p(\bx_c)>0$. This establishes
the constant. 

Since $m_p(\cdot)$ is continuous and bounded, there always exists a
finite $M_1>0$ such that $|m_p(\bx)|<M_1$ for all $\bx\in\Re^d$.
Therefore, $\max_{\bx\in\Re^d}|m_p(\bx)-m_p(\bX_{\mathbf{m}})|<2M_1$.
Let $\tau_m=\mbox{e} ^{m^{1/4}}$. Since $2M_1/\tau_m\rightarrow 0$ as
$m\rightarrow\infty$, Condition~(\ref{Condition}) of Theorem~3.8 of
\cite{Shao_1995} holds. Thus, the bootstrap variance estimator is
strongly consistent:
$\lim_{m\rightarrow\infty}\lim_{B\rightarrow\infty}
m\widehat{\sigma}_{I}^2=\sigma_{\mu}^2$ almost surely. 

Next, we will show
$\lim_{m\rightarrow\infty}m\sigma_I^2=\sigma_{\mu}^2$ by proving a
multi-variate version of Theorem~1.1 in \cite{Lehmann_Casella_1998}, 
Chapter~6. Let $L(\bX_{\mathbf{m}},\bx_c)$ denote the line segment
joining $\bX_{\mathbf{m}}$ and $\bx_c$. By the Multivariate Taylor
Formula \citep{Serfling_2002}, 
\[ 
m_p(\bX_{\mathbf{m}})=m_p(\bx_c)+\nabla 
m_p(\bx_c)^\top
(\bX_{\mathbf{m}}-\bx_c)+\frac{1}{2}(\bX_{\mathbf{m}}-\bx_c)^\top 
\nabla
^2m_p(\bx_c)(\bX_{\mathbf{m}}-\bx_c)+\mathcal{R}(\bX_{\mathbf{m}},\bx_c).
\]
The remainder term 
\[
\mathcal{R}(\bX_{\mathbf{m}},\bx_c)=
\frac{1}{3!}\sum_{i_1=1}^d\sum_{i_2=1}^d\sum_{i_3=1}^d
\left. \frac{\partial^3 
m_p(x_1,\ldots,x_d)}{\partial x_{i_1}\partial x_{i_2}
\partial
x_{i_3}} \right|_{\bx=\bz}\prod_{j=1}^3(X_{\mathbf{m},i_j}-x_{c,i_j})  
\]
where $\bz$ denotes a value in the interior of
$L(\bX_{\mathbf{m}},\bx_c)$, and $X_{\mathbf{m},i}$ and $x_{c,i}$
denote the $i$th components of the vectors $\bX_{\mathbf{m}}$ and
$\bx_c$.  By taking the expectation over both sides, we have 
\begin{equation}
\label{eq.Theorem3_1}
\mbox{E}[m_p(\bX_{\mathbf{m}})]=m_p(\bx_c)+\frac{1}{2}\mbox{E}
\left[
(\bX_{\mathbf{m}}-\bx_c)^\top\nabla^2
m_p(\bx_c)(\bX_{\mathbf{m}}-\bx_c) \right]+ 
\mbox{E}[\mathcal{R}(\mathbf{X}_{\mathbf{m}},\bx_c)] 
\end{equation}
where $\nabla^2$ is the Hessian operator.

We will show that the second and third terms on the RHS of
Equation~(\ref{eq.Theorem3_1}) are $O(m^{-1})$ and $O(m^{-2})$,
respectively, under Assumption~5.  Since all of the input processes
are independent, we establish the result for one input distribution
$F^c$ without loss of generality. 

We prove the result for $\bx_c$ being the generic $h$th-order moment,
$x_{c,h}=\mbox{E}(Z_1^h)<\infty$, and $X_{m,h}=
m^{-1}\sum_{j=1}^mZ_j^h$ for $Z_j\stackrel{\mbox{iid}}\sim F^c$.  

Let $C_{ij}\equiv\frac{1}{2}[\nabla^2 m_p(\bx_c)]_{i,j}$. We
first consider components of the second term on the RHS
of Equation~(\ref{eq.Theorem3_1}).
\begin{eqnarray}
\lefteqn{ \mbox{E}[C_{ij}(X_{m,i}-x_{c,i})(X_{m,j}-x_{c,j})]
=C_{ij}\mbox{E}\left[\frac{1}{m}\sum_{k_1=1}^m(Z_{k_1}^i-x_{c,i})\cdot\frac{1}{m}
\sum_{k_2=1}^m(Z_{k_2}^j-x_{c,j})\right] } \nonumber \\
&=& \frac{C_{ij}}{m^2}\mbox{E}\left[\sum_{k=1}^m(Z_k^i-x_{c,i})(Z_k^j-x_{c,j})
+\sum_{k_1\neq k_2}(Z_{k_1}^i-x_{c,i})(Z_{k_2}^j-x_{c,j})\right] \nonumber  \\
&=& \frac{C_{ij}}{m^2}\mbox{E}\left[\sum_{k=1}^m(Z_k^i-x_{c,i})(Z_k^j-x_{c,j})
+0\right]=O\left(\frac{1}{m}\right) \nonumber. 
\end{eqnarray}
The last two steps follow because the $Z_i$ are i.i.d.\ and Assumption~5
holds.  Thus, the second term on the RHS of
Equation~(\ref{eq.Theorem3_1}) is
\[
\frac{1}{2}\mbox{E}
[(\bX_{\mathbf{m}}-\bx_c)^\top\nabla^2
m_p(\bx_c)(\bX_{\mathbf{m}}-\bx_c)] = \sum_{i=1}^d\sum_{j=1}^d
\mbox{E}[C_{ij}(X_{m,i}-x_{c,i})(X_{m,j}-x_{c,j})] =
O\left(\frac{1}{m}\right). 
\]
Similarly, for the components of the third term of the RHS of
Equation~(\ref{eq.Theorem3_1}), we have 
\begin{eqnarray}
\lefteqn{ D_{ijk}\mbox{E}[(X_{m,i}-x_{c,i})(X_{m,j}-x_{c,j})(X_{m,k}-x_{c,k})] } \nonumber \\
&=& D_{ijk}\mbox{E}\left[\frac{1}{m^3}\sum_{k_1=1}^m(Z_{k_1}^i-x_{c,i})
\cdot\sum_{k_2=1}^m(Z_{k_2}^j-x_{c,j})\cdot\sum_{k_3=1}^m(Z_{k_3}^k-x_{c,k})\right] \nonumber \\
&=& \frac{D_{ijk}}{m^3}\mbox{E}\left[\sum_{k_1=1}^m(Z_{k_1}^i-x_{c,i})(Z_{k_1}^j-x_{c,j})
(Z_{k_1}^k-x_{c,j})+0\right] = O\left(\frac{1}{m^2}\right). \nonumber 
\end{eqnarray}
where
\[
D_{ijk}\equiv \left. \frac{1}{3!}\frac{\partial^3
m_p(x_1,\ldots,x_d)}{\partial x_{i}\partial x_{j}\partial
x_{k}} \right|_{\bx=\bz}.
\]
Again, the last two steps follow because the $Z_i$ are
i.i.d.\ and Assumption~5 holds.  Thus, the third term in
Equation~(\ref{eq.Theorem3_1}) is
\[
\mbox{E}[\mathcal{R}(\mathbf{X}_{\mathbf{m}},\bx_c)]=
\sum_{i=1}^d\sum_{j=1}^d\sum_{k=1}^d
D_{ijk}\mbox{E}[(X_{m,i}-x_{c,i})(X_{m,j}-x_{c,j})(X_{m,k}-x_{c,k})]
=O\left(\frac{1}{m^2}\right). 
\]
Squaring both sides of Equation~(\ref{eq.Theorem3_1}), we have
\begin{equation}
[\mbox{E}(m_p(\bX_{\mathbf{m}}))]^2=m_p^2(\bx_c)+m_p(\bx_c)
\mbox{E}[(\bX_{\mathbf{m}}-\bx_c)^\top\nabla^2 m_p(\bx_c)
(\bX_{\mathbf{m}}-\bx_c)]
+O\left(\frac{1}{m^2}\right) .
\end{equation}

By repeating the same derivation that results in
Equation~(\ref{eq.Theorem3_1}) but using $m_p^2(\cdot)$ instead of
$m_p(\cdot)$, we obtain 
\begin{eqnarray}
\lefteqn{\mbox{E}[m_p^2(\bX_{\mathbf{m}})]=m_p^2(\bx_c)+\frac{1}{2}
\mbox{E}[(\bX_{\mathbf{m}}-\bx_c)^\top\nabla^2m_p^2(\bx_c)(\bX_{\mathbf{m}}-\bx_c)]
+O\left(\frac{1}{m^2}\right) }\nonumber \\
\label{eq.Theorem3_2}
&=& m_p^2(\bx_c)+\mbox{E}\Big[(\bX_{\mathbf{m}}-\bx_c)^\top\nabla m_p(\bx_c)\nabla m_p(\bx_c)^\top (\bX_{\mathbf{m}}-\bx_c) \nonumber \\
&& +(\bX_{\mathbf{m}}-\bx_c)^\top m_p(\bx_c)\nabla ^2m_p(\bx_c)(\bX_{\mathbf{m}}-\bx_c)\Big]+O\left(\frac{1}{m^2}\right). 
\end{eqnarray}
Then,
\begin{eqnarray}
\lefteqn{\mbox{Var}[m_p(\bX_{\mathbf{m}})]=\mbox{E}[m_p^2(\bX_{\mathbf{m}})]
-\Big(\mbox{E}[m_p(\bX_{\mathbf{m}})]\Big)^2 }\nonumber \\
&=&\mbox{E}[(\bX_{\mathbf{m}}-\bx_c)^\top\nabla m_p(\bx_c)
\nabla m_p(\bx_c)^\top(\bX_{\mathbf{m}}-\bx_c)] + O\left(\frac{1}{m^2}\right) \nonumber \\
\label{eq.Th3_mid1}
&=& \mbox{E}[\nabla m_p(\bx_c)^\top(\bX_{\mathbf{m}}-\bx_c)
(\bX_{\mathbf{m}}-\bx_c)^\top\nabla m_p(\bx_c)]+O\left(\frac{1}{m^2}\right) \\
&=& \frac{1}{m}\nabla m_p(\bx_c)^\top \Lambda\nabla m_p(\bx_c)+O\left(\frac{1}{m^2}\right). \nonumber
\end{eqnarray}
Step~(\ref{eq.Th3_mid1}) follows because $\nabla m_p(\bx_c)^\top(\bX_{\mathbf{m}}-\bx_c)$ is a scalar.
Thus, we have $\lim_{m\rightarrow\infty}m\sigma_I^2=\sigma_{\mu}^2$. \blot

\vspace{12pt}

\noindent \textbf{Remark:} The independent variables in our stochastic
kriging metamodel consist of central moments and standardized central
moments, rather than raw moments. However, Theorem~3 can easily be
extended to central and standardized central moments as follows.

Since standardized moments are continuous functions of raw moments,
denoted generically as $g(\cdot)$, we can consider the composite
function $(m_p\circ g)(\cdot)$ and follow steps analogous to those in
the proof of Theorem~3.  Up to the third derivatives we have
\begin{eqnarray*}
(m_p\circ g)^\prime(t) &=& m_p^\prime(g(t))g^\prime(t) \\
(m_p\circ g)^{\prime\prime}(t) &=&
m_p^{\prime\prime}(g(t))[g^\prime(t)]^2+m_p^\prime(g(t))
g^{\prime\prime}(t) \\
(m_p\circ g)^{(3)}(t) &=& 
m_p^{(3)}(g(t))[g^\prime(t)]^3+2m_p^{\prime\prime}(g(t))g^\prime(t)g^{\prime\prime}(t)
+m_p^{\prime\prime}(g(t))g^\prime(t)g^{\prime\prime}(t)+m_p^\prime(g(t))g^{(3)}(t). 
\end{eqnarray*}
Let $u$ denote the
mean, $u_i^\prime$ denote the $i$th order raw moment and $u_i$ denote the
$i$th order central moment. Then the first three central moments can
be expressed as functions of raw moments as follows:
\begin{eqnarray*}
u_1 &=& u, \\
u_2 &=& u_2^\prime-u^2 \\
u_3 &=& u_3^\prime-3uu_2^\prime+2u^3.
\end{eqnarray*}
The first three standardized central moments are $u_1,\sqrt{u_2}$ and
$u_3/u_2^{3/2}$. For a non-degenerate distribution, the second central
moment is positive and bounded away from $0$. Thus, the first three
derivatives $g^\prime,g^{\prime\prime},g^{(3)}$ exist and are finite.

%We can first apply the central limit theorem to the raw moments and
%then apply the multivariate delta method to the composition of
%functions $g(\cdot)\equiv m_p(\mathbf{f}(\cdot))$ with
%$g^\prime(\bx)=m_p^\prime(\mathbf{f}(\bx))\mathbf{f}^\prime(\bx)\neq \mathbf{0}_{d\times
%1}$ almost surely. Thus, by following the similar proof used for raw
%moments, we have $\lim_{m\rightarrow\infty}m\sigma^2_I
%=\lim_{m\rightarrow\infty} \lim_{B\rightarrow\infty}
%m\widehat{\sigma}^2_I=\sigma^2_\mu$, where $\sigma^2_\mu$ is a positive constant.

\subsection{Experiment Design}

To fit SK metamodels we recommend the experiment design developed in
\cite{barton_nelson_xie_2011} which demonstrated robust performance
over a number of test examples. In this section, we briefly review the
basic methodology; for detailed information please refer to
\cite{barton_nelson_xie_2011}.  

The experiment design is not specified a priori; instead the design
space, denoted by $\mathcal{D}$, depends on the real-world data
$\mathbf{z}_{\mathbf{m}}^{(0)}$ that will eventually be resampled. In
this way the design is adaptive.

At a high level, this is the approach: Generate a large number of
bootstrap samples from the real-world data
$\mathbf{z}_{\mathbf{m}}^{(0)}$ and compute the corresponding sample
moments. Find a regular region that encompasses a large fraction of
this sample; this will be the design space. Generate additional
bootstrap samples to test that the regular region does indeed cover
the desired fraction of the feasible space of sample moments, and
refine if necessary. Once satisfied, embed a space-filling design in
the regular region. These design points correspond to input
distribution moments at which to run simulation experiments to fit the
SK metamodel. We provide some more details below. 

Suppose we are interested in a $(1-\alpha)100\%$ CI; we set
$\alpha=0.05$ in our empirical study. We want the experiment
design to lead to a metamodel that is accurate for moments $\bx$ that are
the most likely bootstrap moment vectors generated from
$\mathbf{z}_{\mathbf{m}}^{(0)}$; by ``likely'' we mean, for instance,
covering $q=99\%>(1-\alpha)100\%=95\%$ of the feasible bootstrap
moments. 

To this end we find an ellipsoid that will contain an
independent bootstrap moment vector obtained by random sampling from
$\mathbf{z}_{\mathbf{m}}^{(0)}$ with probability at least $q$. We then
generate a space-filling experiment design inside this ellipsoid.  The
 procedure for constructing the design is as follows:
\begin{enumerate}

\item Generate $B_0$ bootstrap resamples from
$\mathbf{z}_{\mathbf{m}}^{(0)}$ and compute the corresponding sample
moments to generate a set of sample moments
$D_T=\{\widehat{\mathbf{X}}_{\mathbf{m}}^{(b)}, b=1,2,\ldots,B_0\}$. 

\item Find the smallest ellipsoid $E$ such that it contains the fraction $q$
of the data in $D_T$ when the ellipsoid's center and shape are the sample
mean and covariance matrix, respectively, of the elements of $D_T$.

\item Perform a hypothesis test where the null hypothesis is that a
bootstrap moment will be contained in this ellipsoid with probability
at least $q$. This requires computing the number of bootstrap moment resamples, denoted
by $B_1$, and the constant $c$ that defines the rejection region to
attain the desired Type I error and power for the test. 

\item Generate $B_{1}$ additional independent bootstrap resamples from
$\mathbf{z}_{\mathbf{m}}^{(0)}$ and compute the moments
$\widehat{\mathbf{X}}_{\mathbf{m}}^{(b)},
b=B_0+1,B_0+2,\ldots,B_0+B_1$.  If more than $c$ of these $B_1$
resamples are contained in the ellipsoid, then accept the current $E$
as the design space. Otherwise, add these bootstrap resamples to
$D_T$, let $B_0 \leftarrow B_0 + B_1$ and go to Step~2 to update the
ellipsoid.

\item Generate $k$ space-filling design points in the ellipsoid $E$.
To place design points into this space, we employ an algorithm due to
\cite{SunFarooq_2002}, \S3.2.1, for generating points uniformly
distributed in an ellipsoid. The algorithm first generates the polar
coordinates of a point uniformly distributed in a hypersphere, then
transforms it to Cartesian coordinates, and finally transforms it
again to a point uniformly distributed in an ellipsoid.  The advantage
of this approach is that each element of the initial polar coordinates
are independently distributed, allowing them to be generated
coordinate by coordinate via their inverse cumulative distribution
function. Rather than use randomly chosen points, however, we begin
with a Latin hypercube sample on $(0, 1)^d$.

\item  Assign $n=N/k$ replications to each design point,
where $N$ denotes total computational budget. Together the
transformed Latin hypercube design points and the number of
replications $n$ define the experiment design $\mathcal{D}$.

\end{enumerate}
In our experiments we set Type I error of the hypothesis test to
$0.005$ and its power to $0.95$ when the true probability is $q = 0.97$.

\color{black}

\subsection{Sensitivity of Inference to SK Parameter Estimation Error}
\label{app:sensitivity}

Since the parameters $(\beta_0, \tau^2, \pmb{\theta}, C)$ are unknown,
we use estimators $(\widehat{\beta}_0, \widehat{\tau}^2,
\widehat{\pmb{\theta}}, \widehat{C})$ to form a SK metamodel. However,
the properties of SK, and in particular Theorems~1--3, have only been
established when at least $(\tau^2, \pmb{\theta}, C)$ are known.
Nevertheless, kriging and SK have been observed to provide robust
inference without accounting for parameter-estimation error provided
we employ an adequate experiment design.  Here we report a
small-scale empirical study that examines parameter sensitivity for
our particular problem: forming an ACI for $\mu(\bx_c)$ and assessing
the relative contribution of input uncertainty.

When we use the plug-in estimator $\widehat{C}$, we get an unbiased SK
predictor $\widehat{m}_p(\mathbf{x})$ and small variance inflation,
based on results in \cite{ankenman_nelson_staum_2010}. Therefore, we
focus on  sensitivity to the parameters
$\pmb{\phi} = (\beta_0, \tau^2,\pmb{\theta})$.  These parameters are estimated by 
maximum likelihood using the log-likelihood function 
\begin{equation}
\ell(\pmb{\phi}) = -\frac{k}{2}\ln(2\pi)-\frac{1}{2}\ln[|\Sigma+C|]
-\frac{1}{2}(\bar{\mathbf{Y}}_{\mathcal{D}}-\beta_0\cdot 1_{k\times 1})^\top
(\Sigma+C)^{-1}(\bar{\mathbf{Y}}_{\mathcal{D}}-\beta_0\cdot 1_{k\times 1}) \nonumber
\label{eq.loglikelihood2}
\end{equation}
where $\Sigma$ is a function of $\tau^2$ and $\pmb{\theta}$.  The only
random variable in the log-likelihood is
$\mathbf{Y}_{\mathcal{D}}$, and the estimation uncertainty of
the MLE $(\widehat{\tau}^2,\widehat{\pmb{\theta}})$ is a complex
function of the sampling distribution of
$\mathbf{Y}_{\mathcal{D}}$ (the contribution to uncertainty due
to $\widehat{\beta}_0$ is tractable). 

To study the sensitivity of our ACI to the estimation error of
$(\widehat{\beta}_0, \widehat{\tau}^2,\widehat{\pmb{\theta}})$, we again use the queueing
network example in Section~6. In each macro-replication, we generate
$m$ real-world observations from each input model, find $k$ design
points and run simulations to obtain $\bar{\mathbf{Y}}_{\mathcal{D}}$
and $\widehat{C}$,
now denoted by $\bar{\mathbf{Y}}_1$ and $\widehat{C}_1$.  Then we compare the performance
of our method under two settings:
\begin{description}

\item[Case 1:] Use $\bar{\mathbf{Y}}_1$ and $\widehat{C}_1$ to compute the MLEs for
$(\widehat{\beta}_{01}, \widehat{\tau}^2_1,\widehat{\pmb{\theta}}_1)$,
use them to build the SK metamodel and construct CI$_{+}$ as before
(Section~5.1).

\item[Case 2:] Using the same experiment design, draw another
independent sample of simulation outputs to obtain
$\bar{\bY}_{\mathcal{D}}$ and $\widehat{C}$, denoted by $\bar{\bY}_2$
and $\widehat{C}_2$, and obtain the corresponding MLEs
$(\widehat{\beta}_{02}, \widehat{\tau}^2_2,\widehat{\pmb{\theta}}_2)$.
Use these estimates along with $\bar{\mathbf{Y}}_1$ and
$\widehat{C}_1$ to build the metamodel and again construct CI$_{+}$. 

\end{description}
Notice that in Case 2 we are obtaining GP parameter estimates from a
sample of data that is independent of the data that forms the
metamodel. 

Table~\ref{table:sideExp} shows the coverage and contribution results
based on 1000 macro-replications. The nearly identical performance of
Cases~1 and~2 demonstrates that our procedure is not sensitive to SK
parameter estimation error if we employ the one-stage space-filling
design used in the paper.

\begin{table}[tbp] \color{black}
\caption{Sensitivity of the ACI to the estimation of
$(\widehat{\tau}^2,\widehat{\pmb{\theta}})$. } \label{table:sideExp}
\begin{center}
\begin{tabular}{|c|c|c|c|c|c|c|} 
\hline
$k=40$, $n=50$ & \multicolumn{3}{c|}{Case 1} & \multicolumn{3}{c|}{Case 2} \\ \cline{2-7}
 &  $m=50$ & $m=500$  & $m=5000$ & $m=50$ & $m=500$  & $m=5000$  \\
 \hline
%Coverage of $\mbox{CI}_0$ & 94.5\%  & 96.4\% & 94.2\%  & 94.6\% & 96.8\% & 93.3\% \\ \hline
Coverage of $\mbox{CI}_{+}$ & 95.6\% & 97.9\% & 96.4\% & 95.5\% & 98.1\% & 95.5\%  \\ \hline
%$\mbox{CI}_0$ Width (mean)  & 326 & 27.6 & 4.3  & 326 & 27.5 & 4 
%\\ \hline 
$\mbox{CI}_{+}$ Width (mean) & 340 & 29.4 & 4.6  & 340  & 29.3  & 4.4 
\\ \hline 
%$\mbox{CI}_0$ Width (SD) & 169 & 18.1 & 0.95  & 169  & 18.1  & 0.9 
%\\ \hline 
$\mbox{CI}_{+}$ Width (SD) & 171  & 18.8 &  0.93  & 170 & 18.8 & 0.91  
\\ \hline 
$\widehat{\sigma}_I/\widehat{\sigma}_T$ & 0.972 & 0.960 & 0.925 & 0.971 & 0.963  & 0.922 
\\ \hline
\end{tabular}
\end{center} \color{black}
\end{table}

\color{black}

\end{document}